\documentclass[aps,prx,twocolumn,floatfix,superscriptaddress,amsmath,amssymb,showpacs,preprintnumbers,reprint]{revtex4-1}
\usepackage{graphicx,color}
\usepackage{dcolumn}
\usepackage{bm}
\usepackage{stmaryrd}
\usepackage{latexsym}
\usepackage{amssymb}
\usepackage{amsfonts}
\usepackage{amsmath}
\usepackage{textcomp}

\begin{document}

\title{Orbital-driven Mottness collapse in 1T-TaS$_{2-x}$Se$_x$ transition metal dichalcogenide}

\author{Shuang Qiao}
\affiliation{State Key Laboratory of Low Dimensional Quantum Physics, Department of Physics, Tsinghua University, Beijing 100084, P.R. China}
\affiliation{Institute for Advanced Studies, Tsinghua University, Beijing 100084, P.R. China}
\author{Xintong Li}
\affiliation{State Key Laboratory of Low Dimensional Quantum Physics, Department of Physics, Tsinghua University, Beijing 100084, P.R. China}
\author{Naizhou Wang}
\affiliation{Hefei National Laboratory for Physical Science at Microscale and Department of Physics, University of Science and Technology of China, Hefei, Anhui 230026, P.R. China}
\author{Wei Ruan}
\affiliation{State Key Laboratory of Low Dimensional Quantum Physics, Department of Physics, Tsinghua University, Beijing 100084, P.R. China}
\author{Cun Ye}
\affiliation{State Key Laboratory of Low Dimensional Quantum Physics, Department of Physics, Tsinghua University, Beijing 100084, P.R. China}
\author{Peng Cai}
\affiliation{State Key Laboratory of Low Dimensional Quantum Physics, Department of Physics, Tsinghua University, Beijing 100084, P.R. China}
\author{Zhenqi Hao}
\affiliation{State Key Laboratory of Low Dimensional Quantum Physics, Department of Physics, Tsinghua University, Beijing 100084, P.R. China}
\author{Hong Yao}
\affiliation{Institute for Advanced Studies, Tsinghua University, Beijing 100084, P.R. China}
\affiliation{Innovation Center of Quantum Matter, Beijing 100084, P.R. China}
\author{Xianhui Chen}
\affiliation{Hefei National Laboratory for Physical Science at Microscale and Department of Physics, University of Science and Technology of China, Hefei, Anhui 230026, P.R. China}
\affiliation{Collaborative Innovation Center of Advanced Microstructures, Nanjing University, Nanjing 210093, China}
\author{Jian Wu}
\email{wu@phys.tsinghua.edu.cn}
\affiliation{State Key Laboratory of Low Dimensional Quantum Physics, Department of Physics, Tsinghua University, Beijing 100084, P.R. China}
\affiliation{Innovation Center of Quantum Matter, Beijing 100084, P.R. China}
\author{Yayu Wang}
\email{yayuwang@tsinghua.edu.cn}
\affiliation{State Key Laboratory of Low Dimensional Quantum Physics, Department of Physics, Tsinghua University, Beijing 100084, P.R. China}
\affiliation{Innovation Center of Quantum Matter, Beijing 100084, P.R. China}
\author{Zheng Liu}
\email{zliu@phys.tsinghua.edu.cn}
\affiliation{Institute for Advanced Studies, Tsinghua University, Beijing 100084, P.R. China}
\affiliation{Innovation Center of Quantum Matter, Beijing 100084, P.R. China}

\date{\today}


\begin{abstract}
The layered transition metal dichalcogenide 1T-TaS$_2$ has been recently found to undergo a Mott-insulator-to-superconductor transition induced by high pressure, charge doping, or isovalent substitution. By combining scanning tunneling microscopy (STM) measurements and first-principles calculations, we investigate the atomic scale electronic structure of 1T-TaS$_2$ Mott insulator and its evolution to the metallic state upon isovalent substitution of S with Se. We identify two distinct types of orbital textures - one localized and the other extended - and demonstrate that the interplay between them is the key factor that determines the electronic structure. Especially, we show that the continuous evolution of the charge gap visualized by STM is due to the immersion of the localized-orbital-induced Hubbard bands into the extended-orbital-spanned Fermi sea, featuring a unique evolution from a Mott gap to a charge-transfer gap. This new mechanism of orbital-driven Mottness collapse revealed here suggests an interesting route for creating novel electronic states and designing future electronic devices.
\end{abstract}

\maketitle

\section{INTRODUCTION}

The vicinity of a Mott insulating phase has constantly been a fertile ground for finding exotic quantum states, most notably the high T$_c$ cuprates and colossal magnetoresistance manganites. The layered transition metal dichalcogenide 1T-TaS$_2$ represents another intriguing example. More interestingly, it has been recently found that 1T-TaS$_2$ undergoes a Mott-insulator-to-superconductor transition induced by high pressure \cite{9}, charge doping \cite{8,10,11,15}, or isovalent substitution \cite{12,13,14}.

The nature of the Mott insulator phase and transition mechanism to the conducting state is still under heated debate. Much of the challenge originates from a series of complicated charge density wave (CDW) orders entwined with the electronics phases \cite{1,2,3}. The Mott insulator phase is believed to be a consequence of the commensurate CDW order \cite{3,4,5,6,7}. Superconductivity emerges as the Mott insulator state is suppressed, and the electronic phase diagram \cite{9,11,13,15} displays many similarities to that of the cuprates.

Different scenarios were proposed to account for the Mott-insulator-to-superconductor transition. For the pressurization case, it was assumed that the superconductivity formed within the metallic interdomain spaces of the nearly commensurate CDW (NCCDW) phase \cite{9}. This scenario is consistent with the phase separation picture of the NCCDW phase \cite{50}. It also naturally explains the continuous extension of superconductivity into the incommensurate CDW (ICDW) phase, where the interdomain spaces grow to the whole sample and the CCDW domains completely disappear. On the other hand, for the chemical doping cases, angle-resolved photoemission measurements did not reveal any signature of phase separation \cite{10}, suggesting instead strong electronic hybridization across domains. The electrons were found to form a global melted Mott state with a single electron pocket around the $\mathrm{\Gamma}$ point, from which superconductivity could emerged. Another perspective on the melted Mott state was recently developed by viewing doped 1T-TaS$_2$ as a disordered Mott insulator \cite{20}, in which the interplay of strong interaction and disorder generates a novel pseudogap metallic state.

Most of the previous explanations were essentially based on a one-band Hubbard model \cite{20,30,31}, and the efforts were made to reproduce the phase transition by tuning some parameters. The problems, however, are that the microscopic meaning of the one band basis is ambiguous, and the dependence of the associated parameters on the material details is unclear. Consequently, it heavily relies on the researcher¡¯s perspective to select the key parameters and the tuning range. This leads to rather controversial interpretations emphasizing on different aspects of the material \cite{9,10,12,14,16,17,18,19,20,21,22}, such as spin-orbit coupling, interlayer coupling, p-d hybridization and disorder.

The aim of the present Article is to establish a concrete foundation that describes the low-energy physics in 1T-TaS$_2$. By combining STM measurements and \emph{ab initio} Wannier function analysis, we single out two distinct types of orbital textures, and formulate a multi-orbital effective Hamiltonian based on these two orbital textures. This construction reveals an orbital-driven origin of the Mott-insulator-to-superconductivity transition in 1T-TaS$_2$. We map out the complete transition process by inhomogeneous Se substitution, and demonstrate an excellent agreement between experiment and theory.  The remainder of the Article is organized as follows. Section II describes the experiment and calculation methods. Section III presents the STM and first-principles data on pristine TaS$_2$, based on which the two types of most relevant orbital textures are identified. Section IV discusses results on Se-substituted samples. By taking advantage of the chemical inhomogeneity in these samples, we map out the complete collapsing process of the Mott phase. Section V discusses the theoretical formalism of the orbital-driven transition in details. Section VI concludes this Article.

\section{EXPERIMENT AND CALCULATION METHODS}

The high-quality single crystals of 1T-TaS$_{2-x}$Se$_x$ (0 $<$ \emph{x} $<$ 2) are synthesized by chemical vapor transport method with iodine as the transport agent. The powder of Ta, S, and Se is weighted stoichiometrically and thoroughly grounded. Then the mixture together with 150 mg I$_2$ is sealed into an evacuated quartz tube and placed into a two-zone furnace with the temperature gradient of 1253 K-1153 K for 1 week. The products are washed with ethanol to remove the iodine at the surface.

The STM experiments are performed with a cryogenic variable temperature ultrahigh vacuum STM system. The 1T-TaS$_{2-x}$Se$_x$ crystal is cleaved in situ at \emph{T} = 77 K, and the measurement is taken at 5 K with an electrochemically etched tungsten tip calibrated on a clean Au (111) surface \cite{34}. The STM topography is taken in the constant current mode, and the \emph{dI/dV} spectra are obtained with a standard lock-in technique with modulation frequency \emph{f} = 423 Hz.

The first-principles calculation based on the density functional theory is performed using the Vienna \emph{Ab initio} Simulation Package \cite{35,36,37,38} with a 280 eV plane-wave basis cutoff. The Perdew-Burke-Ernzerh generalized gradient approximation \cite{39} and the projector augmented wave method \cite{40} are employed. The Wannier function analysis is performed using the WANNIER90 code \cite{27,28,29} on the plain DFT level in order to quantify the single-electron properties, such as p-d hybridization and interatomic hopping. The $+$\emph{U} correction is employed to compare with the STM \emph{dI/dV} spectra, which captures the Coulomb interaction of Ta 5d orbitals on the mean-field level, following the simplified (rotational invariant) approach introduced by Dudarev \cite{41,42}. We employ the lattice parameters determined by experiment \cite{43}, and relax the atomic coordination self-consistently until the forces are less than 0.002 eV/$\AA$. To simulate the CCDW order, we employ a  $\sqrt {13} \times \sqrt {13}$ supercell consisting of 13 Ta atoms forming a Star-of-David (SD) structure. The integration over the Brillouin zone is obtained on a $\Gamma$-centered 6 $\times$ 6 $\times$ 1 k mesh. The electronic self-consistent iterations are converged to $10^{-5}$ eV precision of the total energy.

It has been shown by earlier studies that at least for band structure effects alone, 1T-TaS$_2$ should always be describable as a quasi-two-dimensional system \cite{3,44}. Indeed, recent calculations \cite{22} emphasizing the inter-layer orbital texture showed that the Mott-insulating ground state cannot be correctly reproduced once an artificial c-axis stacking sequence is imposed. On the other hand, in 1T-TaSe$_2$ inter-layer coupling plays a more important role \cite{45}. Therefore, we will restrict our discussion within the S-riched regime, and focus on the \emph{intralayer} CCDW order and orbital textures. All our calculations are based on a 2D structural configuration with a 15 $\AA$-thick vacuum space to eliminate coupling between different layers.

\section{ORBITAL TEXTURES IN PRISTINE $TaS_2$}
\subsection{Structural and electronic properties}

\begin{figure}[ht]
\includegraphics[width=0.45\textwidth]{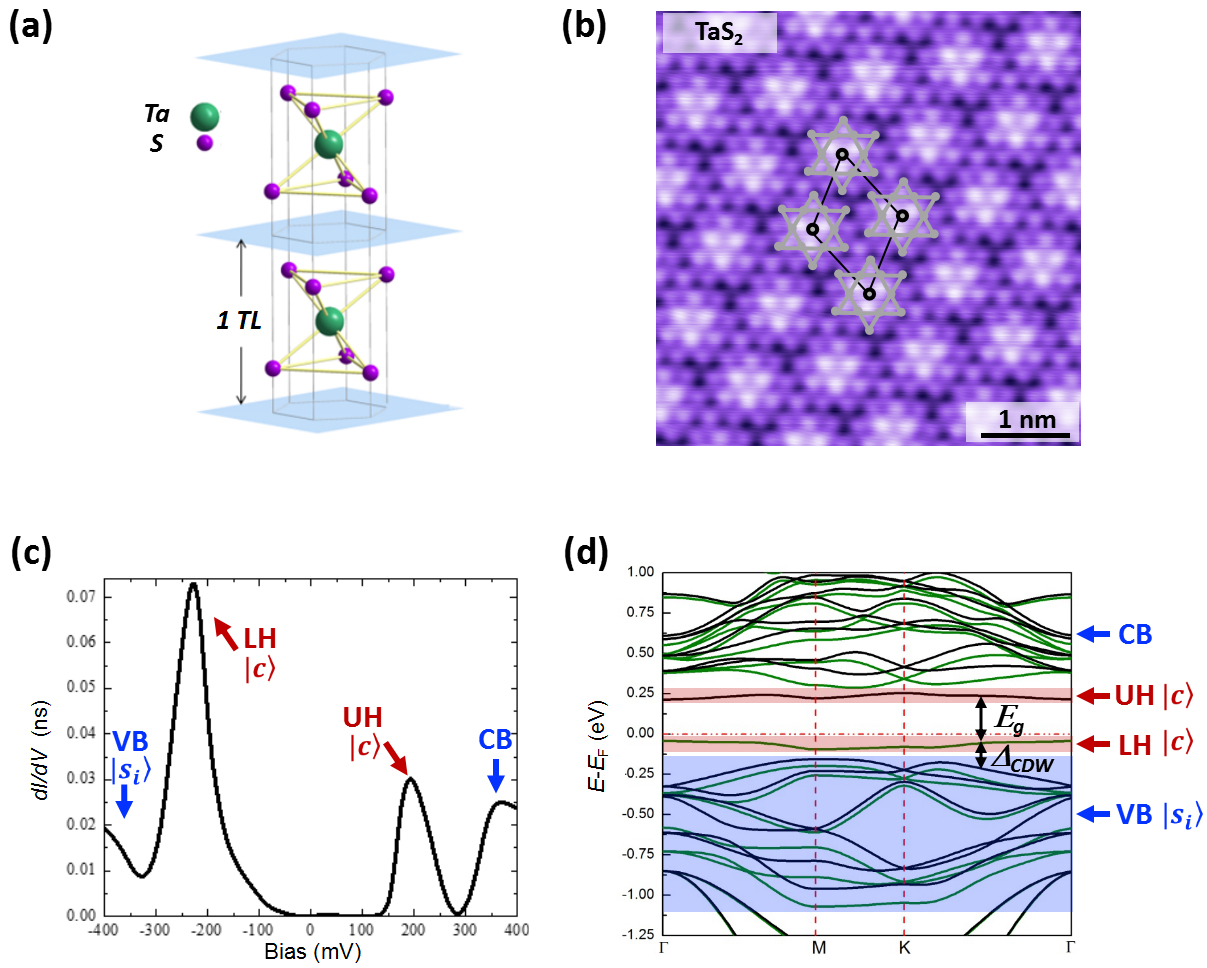}
\caption{\label{fig:structure}  Structural and electronic properties of 1T-TaS$_2$. (a) Crystal structure of 1T-TaS$_2$. (b) Atomic resolved topographic image of 1T-TaS$_2$ with bias voltage \emph{V}$_b$ = 1 V and tunneling current \emph{I}$_t$ = 20 pA. Each bright site on the surface corresponds to a S-atom in the topmost layer. The gray dots and the black lines mark the SD structure and the CCDW superlattice, respectively. The crystal cleaves easily between two adjacent S layers, as indicated by the blue planes in (a), exposing the triangular lattice of S atoms. (c) Spatial averaged STM \emph{dI/dV} spectrum acquired at 5 K. (d) DFT+\emph{U} band structure of 1T-TaS$_2$. The black/green colored curves represent the two spin components. The shaded regions indicate the energy range belonging to the different orbitals.}
\label{1}
\end{figure}

1T-TaS$_2$ has a layered structure as shown in Fig. \ref{1}(a), in which the Ta atoms form a planar triangular lattice sandwiched by two S-atom planes. STM is an ideal experimental technique for probing the atomic scale electronic structure of such a quasi-2D material. Figure \ref{1}(b) displays the STM topographic image of pristine 1T-TaS$_2$, which clearly reveals the CCDW order with a $\sqrt {13} \times \sqrt {13}$ superlattice where 13 Ta atoms form a Star-of-David (SD) structure. Note that the bright spots correspond to the S atoms lying in the uppermost layer, and the position of the Ta atoms can be determined according to the octahedral coordination. In Fig. \ref{1}(c) we show the spatially averaged differential conductance \emph{dI/dV}, which is approximately proportional to the electron density of states (DOS), of pristine 1T-TaS$_2$. An energy gap around the Fermi level (\emph{E}$_F$) can be clearly observed, which is attributed to the Mott-Hubbard gap of the insulating CCDW phase.

These spectroscopic features can be captured by density functional theory (DFT) calculations including the onsite Hubbard \emph{U} correction in a Hartree-Fock manner \cite{23}. As shown in Fig. \ref{1}(d), there is a full charge gap around the (\emph{E}$_F$), which is bounded by two narrow bands, corresponding to the upper and lower Hubbard bands (UHB/LHB), respectively. The gap size is found to depend on the Hubbard repulsion \emph{U} and a choice of \emph{U} = 2.27 eV as previously derived from the linear-response calculation is used to reproduce the experimental value \cite{23}. There is another energy gap below the LHB isolating it from the underneath dispersive valence band continuum [VB shaded in blue in Fig. \ref{1}(d)], which is commonly referred to as the CDW gap ($\Delta$$_{CDW}$) \cite{24,25,26}. This gap is manifested in the \emph{dI/dV} spectrum as a dip below the LHB [Fig. \ref{1}(c)].

\subsection{Low-energy orbital textures and their origin}

\begin{figure*}[ht]
\includegraphics[width=0.6\textwidth]{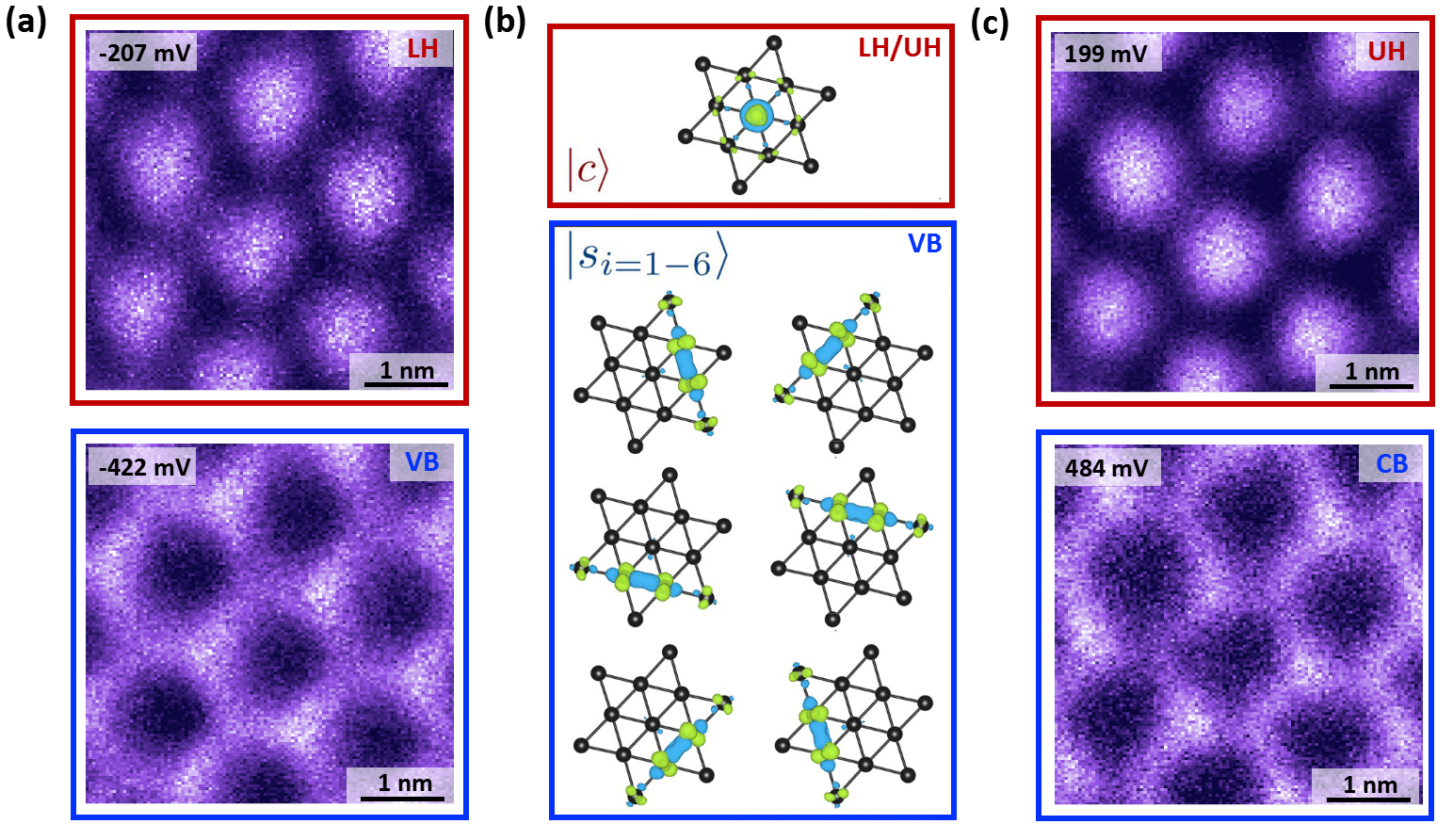}
\caption{\label{fig:structure} Two types of orbital textures in 1T-TaS$_2$. (a)/(c) Differential conductance (\emph{dI/dV}) maps (42\texttimes 42 $\AA^2$) measured at \emph{V}$_b$ = -207/199 mV (up) and \emph{V}$_b$ = -422/484 mV (bottom) with tunneling current \emph{I}$_t$ = 20 pA. The center of each CDW cluster is brighter at -0.2/0.2 V while the edges brighter at -0.4/0.4 V. (b) Isovalue surface of the two types of Wannier functions. }
\label{2}
\end{figure*}

Besides the CCDW superstructure and energy spectrum, equally important is the local DOS (LDOS) mapping at selected energies, which contains the spatial characteristics of electronic wavefunctions, i.e., the orbital texture. Figure \ref{2}(a) displays the \emph{dI/dV}, or LDOS, mapping measured at two representative energies. At \emph{V}$_b$ = -207 mV (upper panel), which corresponds to the LHB, the LDOS map shows a periodic pattern that is commensurate with the CCDW superlattice. By comparing the LDOS map with the topographic image of the same area [Fig. \ref{1}(b)], the bright spots are identified to be the center of SDs. Moving to lower energy with bias \emph{V}$_b$ = -422 mV (lower panel), a distinctly different orbital texture is revealed for the VB: the LDOS map exhibits a kagome-like superstructure. Its bright/dark spots are inverse of the previous texture, indicating contributions mainly from the surrounding Ta orbitals. LDOS maps at positive biases give a nearly symmetric picture, demonstrating the same two types of orbital textures: the UHB \emph{V}$_b$ = 199 mV [Fig. \ref{2}(c) upper panel] corresponds to an orbital texture concentrated on the center of SDs, whereas at higher energy \emph{V}$_b$ = 484 mV [Fig. \ref{2}(c) lower panel] for the conduction bands (CB), the bright spots shift to the surrounding Ta orbitals.

To determine the nature of these two orbital textures discussed above, we proceed by performing \emph{ab initio} Wannier function analysis. Consider that the oxidation state of Ta atom is +4, there is one unpaired 5d electron left per Ta that plays the most important role around the Fermi level. For a SD consisting of 13 Ta atoms, we need seven Wannier functions (without the spin degree of freedom) to fully accommodate these 13 unpaired 5d electrons.  Accordingly, the Bloch states of the seven highest occupied energy bands are chosen to construct the Wannier functions.   The spatial distributions of the resulting maximally localized Wannier functions are plotted in Fig. \ref{2}(b), which automatically separate into two groups. The one denoted by $|c\rangle$ concentrates at the central Ta atom with a typical $d_{z^2}$ geometry. The remaining six Wannier functions distribute dominantly along the edge of the SD ($|s_{\alpha=1...6}\rangle$), which can be viewed as the bonding orbitals of the surrounding Ta atoms. This Wannier function analysis gives a natural explanation for the two types of orbital textures observed by STM. In particular, it clarifies that the kagome-like texture arises from a set of linear-shaped molecular orbitals.

We note that previous STM mapping indicated additional center-surrounding variation of the orbital texture for the deeper occupied states \cite{21}. This feature is also observed in our experiment and calculation. These deep occupied states are associated with a different set of Wannier functions orthogonal to $|c\rangle$ and $|s_{\alpha=1...6}\rangle$, which can be clarified using the same technique. However, those Wannier functions are irrelevant to the low-energy physics. By mapping out the low-energy orbitals, we also note that the previously employed one-band Hubbard model is implicitly based on the $|c\rangle$ orbital alone. Since this orbital is largely localized, it is susceptible to a Mott transition, giving rise to the UHB and LHB. However, it is important to notice that the validity of the one-band Hubbard model is guaranteed only by a large $\Delta$$_{CDW}$. When $\Delta$$_{CDW}$ is reduced, a one-band-to-multi-band transition has to be considered, which in turn plays an important role in the collapse of the Mott phase.

\section{MOTTNESS COLLAPSE UPON $Se$ SUBSTITUTION}

To test the validity of the two-orbital-texture model, we further investigate the electronic structure of 1T-TaS$_2$ under external influence. A rather gentle perturbation to the 1T-TaS$_2$ structure is to modify the buckling of the S atoms around the central Ta, which is expected to affect the CCDW order. Experimentally this can be effectively realized through isovalent substitution of S by Se \cite{12,13}, which has a larger ionic size and is expected to increase the buckling of the surrounding S atoms. We have synthesized a series of Se-doped 1T-TaS$_2$ single crystals, and use these samples to map out the collapse of the Mott phase.

\subsection{Phase transitions}

\begin{figure}[ht]
\includegraphics[width=0.45\textwidth]{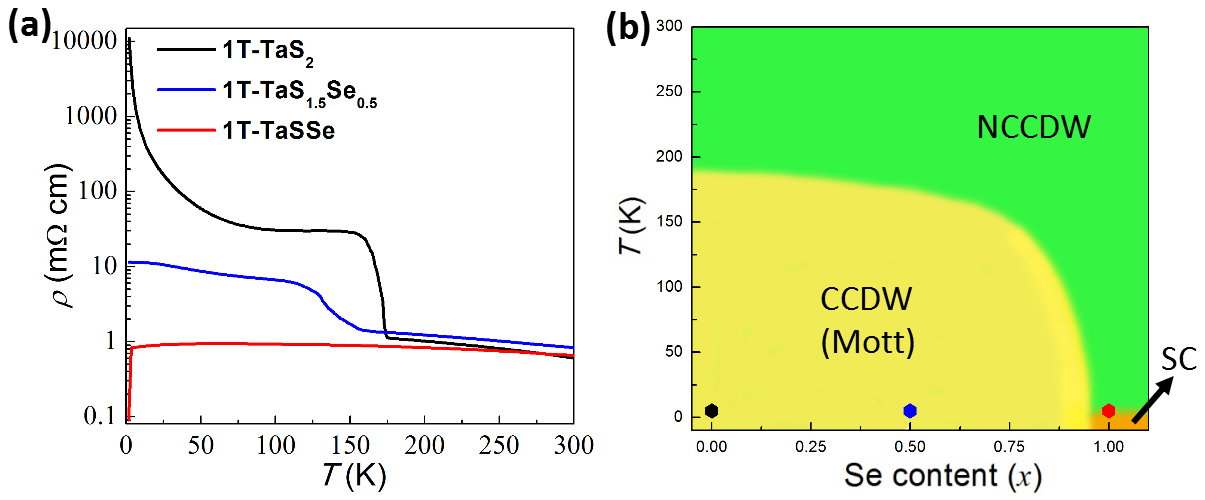}
\caption{\label{fig:structure}  (a) Temperature dependence of the in-plane resistivity of three typical 1T-TaS$_{2-x}$Se$_x$ samples. (b) Schematic electronic phase diagram of 1T-TaS$_{2-x}$Se$_x$ and the location of the three samples under STM measurements.}
\label{3}
\end{figure}

Substituting S with Se is known to suppress the Mott-insulating phase in 1T-TaS$_2$ \cite{12,13}, which can be seen from the transport data. Figure \ref{3}(a) shows the in-plane resistance data of three 1T-TaS$_{2-x}$Se$_x$ samples that are used for discussion in this work. For the pristine 1T-TaS$_2$, there is a well-defined CCDW ordering around 180 K. Below this transition temperature, the resistance exhibits a typical insulating behavior. Clearly, Se substitution strongly suppresses this phase transition. For the \emph{x} = 0.5 sample, the resistance jump around the phase transition point is weakened. For the \emph{x} = 1 sample, it completely vanishes, and a flat resistance curve stays unperturbed until about 3 K, at which a sudden drop occurs. This is known to be a superconducting transition. Figure \ref{3}(b) displays the schematic phase diagram \cite{13}, and the location of the three samples under discussion.

We note that for Se-substitution, due to its chemical similarity to S, the disorder effect is expected to be weak. This can be observed from the transport data: above the CCDW transition temperature, the resistance curves of the three samples almost coincide; the difference only becomes evidence after the CCDW order kicks in.

\subsection{Topographic change}

\begin{figure*}[ht]
\includegraphics[width=0.756\textwidth]{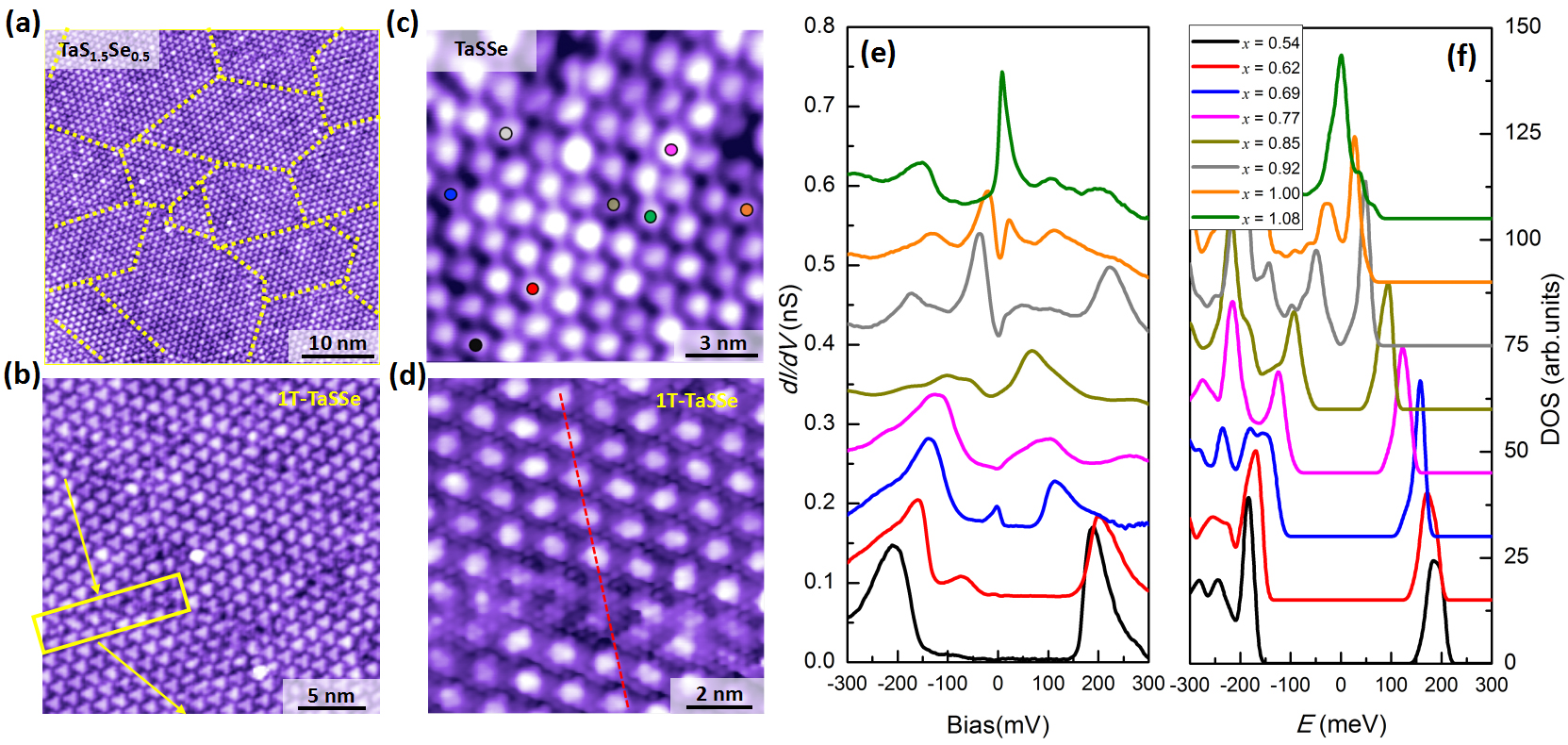}
\caption{\label{fig:structure} Structural and electronic properties evolution of Se substituted 1T-TaS$_2$. Topographic images acquired with \emph{V}$_b$ = -500 mV and \emph{I}$_t$ = 20 pA on the 1T-TaS$_{2-x}$Se$_x$ samples with the averaged Se concentration (a) \emph{x} = 0.5 and (b)$-$(d) \emph{x} = 1. (e) Local STM \emph{dI/dV} spectroscopy measured at various locations on the surface as marked in (c). (f) DFT+\emph{U} density of states of Se-substituted SDs in comparison with (e). In (a), the yellow dot lines indicate the sharp domain boundaries. In (b) and (d), the marks guide for eyes the rotational and translational mismatch in different domains.}
\label{4}
\end{figure*}

How does Se-substitution modify the CCDW order? Figure \ref{4}(a) displays a large area STM image on the surface of the \emph{x} = 0.5 sample. Unlike the pristine sample, the surface structure now splits into domains with varied domain size and sharp boundaries in between [Fig. \ref{4}(a)], which resembles the so-called \textquotedblleft mosaic\textquotedblright phase in 1T-TaS$_2$ induced by voltage pulse in STM experiments \cite{32,33}. Increasing the Se concentration from \emph{x} = 0.5 to \emph{x} = 1 creates more domain walls and defects, but the general topographic properties remain the same. A closer examination around the domain walls reveals both 30\textdegree degree rotational and translational mismatch between different domains [Figs. \ref{4}(b) and (d)]. According to the transport data, this domained phase in the \emph{x} = 1 sample  should adiabatically extend to the high-\emph{T} NCCDW regime, in which the domains might shrink in size and the sharp domain boundaries melt into finite-width interdomain channels free of CCDW order \cite{50}. However, close to the superconducting transition temperature, our STM image shows that the CCDW domains fill the whole space. Within a single domain, the CCDW superlattice retains the SD patterns as that in the undoped TaS$_2$ sample.

\subsection{Electronic inhomogeneity}

STM \emph{dI/dV} spectra are measured at various locations on the surface of the \emph{x} = 1 sample [Fig. \ref{4}(c)], which are found to be surprisingly random. By ordering them with respect to the gap size, we observe a complete evolution [Fig. \ref{4}(e)]. There are still areas (small in size) showing a spectrum similar to that in pristine 1T-TaS$_2$ with a nearly unperturbed Mott gap (black curve). Moving upwards, the Mottness gap shrinks continuously. With further shrinking of the Mott gap, a broad V-shaped gap forms near \emph{E}$_F$, which looks reminiscent of the pseudogap in underdoped cuprates. The V-shaped gap also shrinks gradually, and a finite DOS appears at \emph{E}$_F$, indicating the transition into a metallic state. Eventually the gap vanishes completely and a sharp peak emerges near \emph{E}$_F$, which looks similar to the Van Hove singularity (VHS) in overdoped cuprates.

A natural speculation on the origin of the inhomogeneous spectra is that the Se concentration in different SDs varies. We have calculated a series of SD structures by progressively replacing the S atoms with Se atoms. For a given Se concentration, the lowest-energy substitution configuration is determined, and the DFT structural relaxation confirms that the main effect of Se substitution is to increase the local buckling of the Ta-Se-Ta bonding geometry. We plot the DOS evolution as a function of Se concentration in Fig. \ref{4}(f), rendering a direct comparison with the STM \emph{dI/dV} spectra. The overall evolution of the gap can be nicely reproduced in the theoretical simulation. Besides, several distinct features, e.g. the U-shaped Mottness gap, the V-shaped pseudogap, and the sharp VHS-type peak, are also in good agreement.

Before explaining the underlying orbital-driven physics, it is worth addressing two possible concerns. Firstly, it is technically difficult to exclude the existence of other sources of inhomogeneity. However, given that our pristine sample is of very high quality, we believe that the predominant effect of Se substitution is to generate chemical inhomogeneity. The degree of inhomogeneity may rely on the sample preparation condition, cf. a previous experiment has reported complete ordering of S/Se atoms in the TaSSe sample \cite{14}. Secondly, our calculation only considers doping one single SD with a 2D periodic boundary condition. We assume that the local electronic properties are primarily governed by the intra-cluster coupling within the SD, and the chemical variation in the nearby SDs do not qualitatively change the local spectrum evolution. However, for the global or spatially averaged properties, such as transport behaviors and photoemission spectrum, the situation is beyond this local approximation, which could be an interesting question for further investigation.

\subsection{Orbital shifts}

\begin{figure}[ht]
\includegraphics[width=0.45\textwidth]{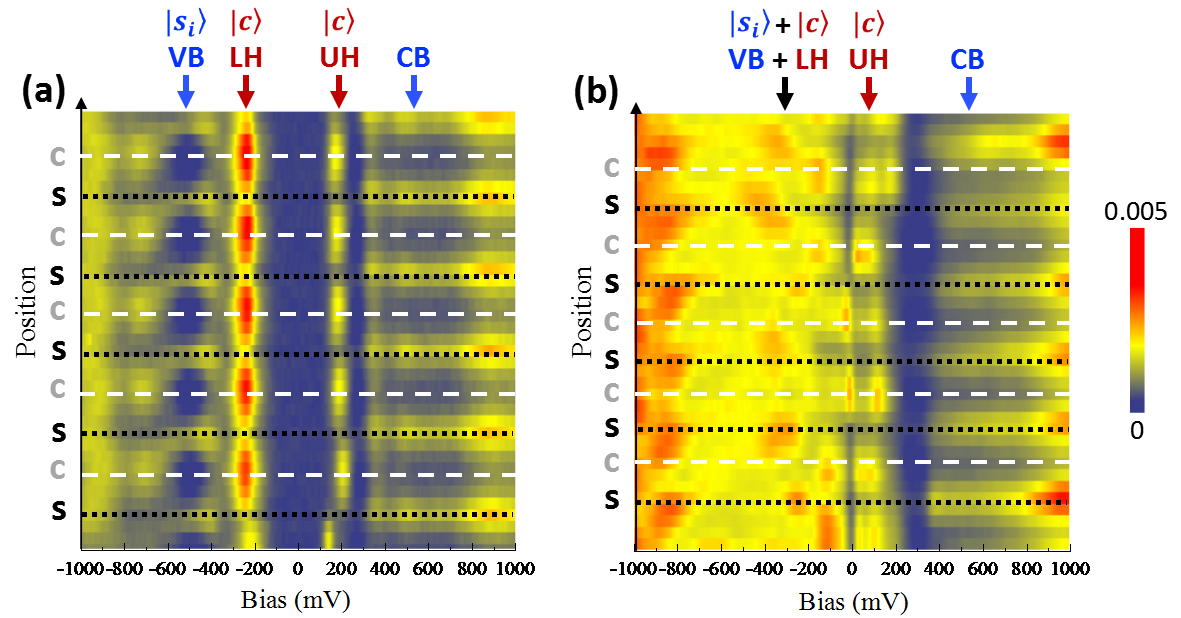}
\caption{\label{fig:structure} STM line map. The \emph{dI/dV} data for (a) \emph{x} = 0 and (b) \emph{x} = 1 case in an extensive two-dimensional parameter space. The horizontal axis is the STM bias and the vertical axis is the position. The white dash (black dot) lines mark the center (surrounding) of the SDs.}
\label{5}
\end{figure}

Due to the spatial inhomogeneity, a 2D periodic pattern as shown in Fig. \ref{2} no longer exists. To uncover the changes of the relevant orbitals, we perform a comparative analysis on the spectra linecut crossing multiple SDs before and after Se-substitution.

In Fig. \ref{5}(a), we first present the linecut map of pristine 1T-TaS$_2$. The horizontal axis is the STM bias, ranging from -1000 meV to 1000 meV. The vertical axis is the position, measured along the line crossing five SDs. The LHB and UHB are evident at the bias of -200 and +200 meV. Along the position axis, they exhibit a periodic intensity variation, which peaks at the center of each SD (c) and weakens at the surrounding (s). The peak of the \emph{dI/dV} shifts by half the period when moving the bias below (above) the LHB (UHB), indicating that the orbital component changes from $|c\rangle$ to $|s\rangle$.

For the doped sample [Fig. \ref{5}(b)], the orbital textures vary from one SD to another, and their spatial distributions become complicated. We first note that the deep CB/VB states, e.g. states around $\pm$1000 meV, remain largely periodic in space, exhibiting the similar textures as in the pristine sample, which again suggests that these states are not involved in the electronic phase transition. Around the Fermi level, the UHB can still be identified, which is bound by two clear gaps in most SDs. Despite a spatially varied magnitude, the UHB in general shifts to the lower energy, leaving a large gap between the CB, and almost touching the VB in some SDs. The strong narrow LHB peak now merges with the top of the VB. For some SDs, a narrow peak around the SD center appears as a remnant of the original LHB.  In general, the surrounding orbitals play an equally important role in the low-energy physics. Because of these orbital shifts, the Mott-insulator-to-metal transition occurs.

\begin{figure*}[ht]
\includegraphics[width=0.738\textwidth]{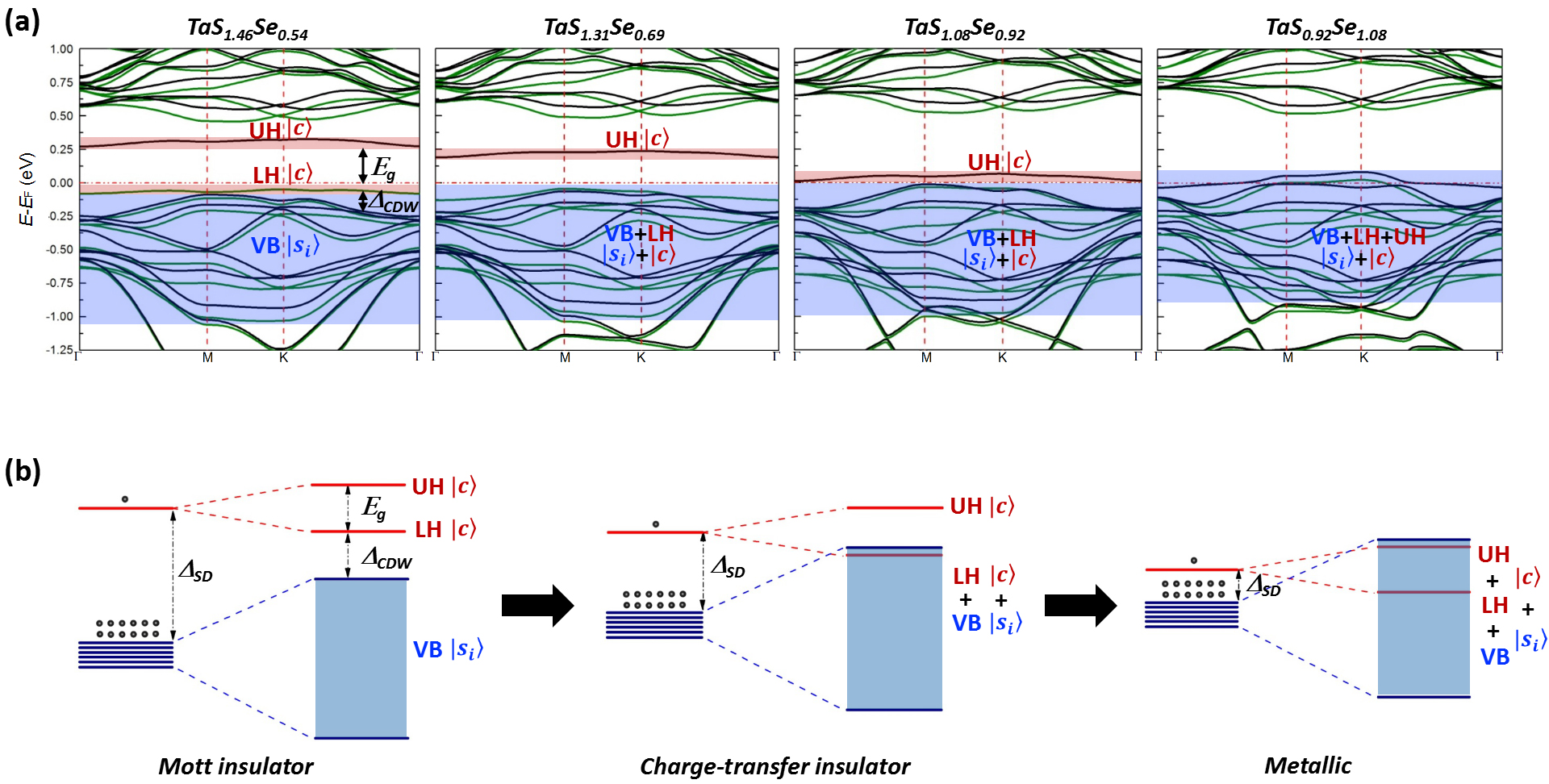}
\caption{\label{fig:structure} (a) DFT+U band structures of Se substituted 1T-TaS$_2$. The meaning of the colors is the same as in Fig. \ref{1}(d). (b) Schematic energy level diagrams of the Mott insulating phase, the charge-transfer insulator phase and the metallic phase.}
\label{6}
\end{figure*}

These orbital shifts can be further compared with the DFT+U band structures. Figure. \ref{6}(a) traces the band evolution as the Se concentration increases.  The results indicate progressive downward shift of UHB and LHB, in agreement with the STM line map. At low-doping concentration, despite a reduction of $\Delta$$_{CDW}$, \emph{E}$_g$ stays unchanged as it is dictated by the effective Hubbard \emph{U} of the $|c\rangle$-orbital alone. After the LHB touches the $|s_{\alpha}\rangle$-orbital bands, the Mott gap transforms into the charge-transfer type gap because now the lowest energy excitation is from the continuum band formed by the $|s_{\alpha}\rangle$-orbitals to the UHB formed by the $|c\rangle$-orbital. The gap size shrinks continuously upon further doping. According to the calculation, the critical point is reached at \emph{x}  $ \approx $ 0.92, where the charge-transfer gap becomes zero. Beyond this critical doping concentration, the UHB also sinks into the $|s_{\alpha}\rangle$-orbital bands, driving the system into the metallic regime. For the high-doping case, the $|c\rangle$-orbital and $|s_{\alpha}\rangle$-orbitals are strongly hybridized so that we can barely see the remnants of the narrow UHB around the $\Gamma$ point at \emph{E}$_F$. The excellent agreement between theory and experiment confirms the validity of the orbital-driven Mott insulator to metal transition, in which the orbital sequence determines the electronic property. We note that in the calculation no additional complexities, such as spin-orbit coupling, interlayer coupling and disorder, are introduced. In addition, after Se substitution, Wannier function analysis shows that the geometry of the $|c\rangle$ and $|s_{\alpha}\rangle$ orbitals remain nearly unchanged. Therefore, we believe that the interplay between the $|c\rangle$-orbital and $|s_{\alpha}\rangle$-orbitals represents a universal origin of Mottness collapse in 1T-TaS$_2$.

\section{THEORETICAL FORMALISM}

To capture the essential physics, we propose a multi-orbital Hubbard model as formulated below:

\begin{eqnarray}
H&=&\Delta_{SD} \sum_I c_I^\dag c_I+t_{cs} \sum_{I\alpha}(c_I^\dag s_{I\alpha}+H.c.) \nonumber \\
&+&\sum_{(I\alpha,J\beta)} t_{ss}^{I\alpha,J\beta}s_{I\alpha}^\dag s_{J\beta}+U_c\sum_I c_{I\uparrow}^\dag c_{I\downarrow}^\dag c_{I\downarrow} c_{I\uparrow},
\end{eqnarray}
where $\Delta$$_{SD}$ is the onsite energy difference between $|c\rangle$ and $|s_{\alpha=1...6}\rangle$ generated by the formation of SD [Fig. \ref{6}(b)], and \emph{I,J} indexes the SDs. The central orbital $|c\rangle$ experiences an effective onsite Coulomb repulsion $U_c$ of the magnitude of LHB/UHB splitting $\sim$0.25 eV, whereas the interaction terms of $|s_{\alpha=1...6}\rangle$ are neglected considering they are less localized. Note that the Wannier function $|c\rangle$ is also a molecular orbital [Fig. \ref{2}(b)], and the associated Coulomb repulsion $U_c$ is much smaller than its original 5d atomic-orbital counterpart as used in the DFT+U calculation. The values of $\Delta$$_{SD}$  and the dominant hopping parameter (\emph{t}) can be extracted by exploiting the same Fourier transformation as used in the Wannier function analysis (Tab. I).

\begin{figure}[ht]
\includegraphics[width=0.38\textwidth]{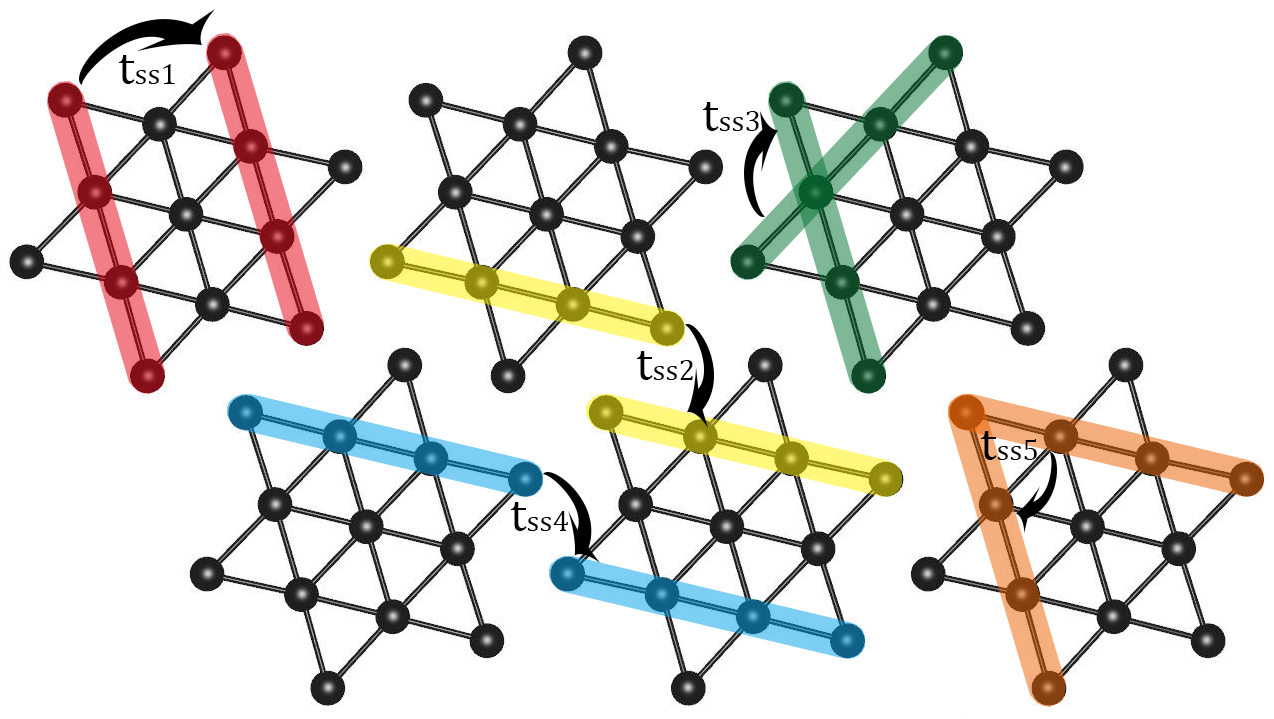}
\caption{Physical meaning of the hopping parameters between the surrounding orbitals listed in Tab. I}
\label{7}
\end{figure}

For pristine 1T-TaS$_2$, the onsite energy of $|c\rangle$ is 0.21 eV higher than $|s_{\alpha=1...6}\rangle$. The largest hopping amplitude is $t_{sc}$ that is between $|c\rangle$ and $|s_{\alpha=1...6}\rangle$, followed by several hopping processes among $|s_{\alpha=1...6}\rangle$ (Fig. \ref{7}). Intercluster hopping is possible via $t_{ss2}$ and $t_{ss4}$. The set of obtained parameters for pristine 1T-TaS$_2$ is susceptible to a Mott transition owing to a relatively large positive $\Delta$$_{SD}$. For the Se-substituted sample, we present in Tab. I the parameters obtained from calculations on the (\emph{x} = 0.92) supercell. Transition from the Mott phase to a metal is in principle described by Eq. (1) when the parameters vary, but solving Eq. (1) including the Hubbard term requires the hard-core many-body calculation that is beyond the scope of the current work. Nevertheless, a quick test can be made by only keeping the largest $t_{ss}$ and $t_{sc}$ terms and then diagonalizing the 7-by-7 Hamiltonian matrix. It shows that the highest eigenstate transforms from a dominant $|c\rangle$ character to a $|s_{\alpha=1...6}\rangle$ character, which reflects the concept of the orbital-driven transition.

\begin{table}
\begin{tabular}{c|c|c|c}
  \hline\hline
  Unit (eV)& Pristine  & Se-doped  & Struct. mod.\\ \hline
  $\Delta_{SD}$ & 0.212 & 0.146 & 0.143\\
  $t_{sc}$ & 0.162 & 0.099 & 0.105\\
  $t_{ss1}$ & 0.150 & 0.175 & 0.171 \\
  $t_{ss2}$ & 0.091 & 0.089 & 0.087 \\
  $t_{ss3}$ & 0.072 & 0.021 & 0.013 \\
  $t_{ss4}$ & 0.050 & 0.047 & 0.043 \\
  $t_{ss5}$ & 0.042 & 0.043 & 0.036 \\
   \hline\hline
\end{tabular}
\caption{Single-electron parameters from Wannier function analysis. The three columns correspond to the pristine 1T-TaS$_2$, 1T-TaS$_{2-x}$Se$_x$ (\emph{x} = 0.92) and an artificial 1T-TaS$_2$ that retains the same structure of 1T-TaS$_{2-x}$Se$_x$,respectively.}
\end{table}

According to this effective model, we schematically summarize the overall physics of the orbital-driven electronic structure evolution in Se doped 1T-TaS$_2$ [Fig. \ref{6}(b)]. The key ingredient is the organization of the two types of orbital textures $|c\rangle$ and $|s_{\alpha=1...6}\rangle$ accompanied by the formation of the SD structures. The central orbital is isolated by the surrounding ones, making inter-cluster hopping of central-orbital electrons possible only via high-order process. It is thus susceptible to a Mott transition when $|c\rangle$ is separated from $|s_{\alpha}\rangle$ in energy. On the other hand, the $|s_{\alpha}\rangle$ orbitals are extended with sizable hopping amplitude, naturally giving rise to a metallic phase when $|c\rangle$ and $|s_{\alpha}\rangle$ mix.

Previous DMFT calculations on the effective one-band Hubbard model \cite{46,47,48}, manually tuned the \emph{U} value, or equivalently, the \emph{t/U} ratio over a very wide range in order to trace the Mott-insulator-to-metal transition. As discussed above, the one-band Hubbard model is a downfolded version of our multi-orbital model by integrating out the $|s_{\alpha=1...6}\rangle$-orbitals. The downfolded \emph{U} is affected not only by $U_c$ associated with the $|c\rangle$-orbital, but also the energy separation between $|c\rangle$ and $|s_{\alpha=1...6}\rangle$. In this sense, our work complements the previous speculation on a change of effective \emph{t/U} ratio that induces the transition, and presents an explicit formalism to explain why the effective \emph{t/U} ratio varies so greatly.

Finally, it is worth mentioning several points that bridges Se substitution with mechanical modulation. As noted in Sec. IVC, the main effect of Se substitution is to increase the local buckling of the Ta-Se-Ta bonding geometry. In calculation, we have tested to retain this buckled geometry, but change Se back to S. It is found that very similar band structure and single-electron parameters can be obtained (c.f. the last two columns of Tab. I). In this sense, the possible structural modification as demonstrated by STM pulsing \cite{32,33} may share the same orbital-driven physics as discussed above.

\section{CONCLUSION}

Our experimental observations and first-principle calculations in 1T-TaS$_2$ reveal a new mechanism of Mottness collapse, which is fundamentally different from previously known examples such as charge-doping-induced collapse in cuprates and pressure-induced collapse in weak Mott insulators. In the layered 1T-TaS$_2$, the isovalent substitution of S with Se does not induce additional charge carriers. Instead, it tunes the CDW order, and in turn drives the direct Mott gap into an effective \textquotedblleft charge-transfer\textquotedblright gap. To the best of our knowledge, it is the first time that such a novel evolution between Mott gap and charge-transfer gap can be adiabatically accessed in the same material. Upon further substitution, after the charge gap fully collapses, the low energy degrees of freedom include the Fermi-surface-forming itinerant carriers and fluctuating local magnetic moments, just like in heavy-fermion systems and iron pnictides/chalcogenides. It is thus interesting to ask whether this parent state for superconductivity is a heavy fermion liquid or a magnetic metal. A even more unique feature of 1T-TaS$_2$ is that the Mott physics is built upon a triangular lattice, in which spin frustrating is intrinsically coded. Therefore, the magnetic ground state of 1T-TaS$_2$ might be highly nontrivial. We believe that the richness of the orbital-tunable electronic structure in Se-doped TaS$_2$ will shed important insights into several important topics of condensed matter physics.

\section{Acknowledgements}

This work is supported by NSFC under grant No 11534007 and MOST under grant No 2015CB921000 of China. Z.L. is supported by Tsinghua University Initiative Scientific Research Program. H. Y. is supported by NSFC under Grant No. 11474175. Z. L. and H. Y also acknowledge supports from the ¡°Thousand Talents Plan of China¡± (for Young Professionals). N.W. and X.C. acknowledge the ¡°Strategic Priority Research Program (B)¡± of the Chinese Academy of Sciences (Grant No. XDB04040100). S. Q and X. L contributed equally to this work.


\begin{thebibliography}{49}%
\makeatletter
\providecommand \@ifxundefined [1]{%
 \@ifx{#1\undefined}
}%
\providecommand \@ifnum [1]{%
 \ifnum #1\expandafter \@firstoftwo
 \else \expandafter \@secondoftwo
 \fi
}%
\providecommand \@ifx [1]{%
 \ifx #1\expandafter \@firstoftwo
 \else \expandafter \@secondoftwo
 \fi
}%
\providecommand \natexlab [1]{#1}%
\providecommand \enquote  [1]{``#1''}%
\providecommand \bibnamefont  [1]{#1}%
\providecommand \bibfnamefont [1]{#1}%
\providecommand \citenamefont [1]{#1}%
\providecommand \href@noop [0]{\@secondoftwo}%
\providecommand \href [0]{\begingroup \@sanitize@url \@href}%
\providecommand \@href[1]{\@@startlink{#1}\@@href}%
\providecommand \@@href[1]{\endgroup#1\@@endlink}%
\providecommand \@sanitize@url [0]{\catcode `\\12\catcode `\$12\catcode
  `\&12\catcode `\#12\catcode `\^12\catcode `\_12\catcode `\%12\relax}%
\providecommand \@@startlink[1]{}%
\providecommand \@@endlink[0]{}%
\providecommand \url  [0]{\begingroup\@sanitize@url \@url }%
\providecommand \@url [1]{\endgroup\@href {#1}{\urlprefix }}%
\providecommand \urlprefix  [0]{URL }%
\providecommand \Eprint [0]{\href }%
\providecommand \doibase [0]{http://dx.doi.org/}%
\providecommand \selectlanguage [0]{\@gobble}%
\providecommand \bibinfo  [0]{\@secondoftwo}%
\providecommand \bibfield  [0]{\@secondoftwo}%
\providecommand \translation [1]{[#1]}%
\providecommand \BibitemOpen [0]{}%
\providecommand \bibitemStop [0]{}%
\providecommand \bibitemNoStop [0]{.\EOS\space}%
\providecommand \EOS [0]{\spacefactor3000\relax}%
\providecommand \BibitemShut  [1]{\csname bibitem#1\endcsname}%
\let\auto@bib@innerbib\@empty
\bibitem [{\citenamefont {Sipos}\ \emph {et~al.}(2008)\citenamefont {Sipos},
  \citenamefont {Kusmartseva}, \citenamefont {Akrap}, \citenamefont {Berger},
  \citenamefont {Forro},\ and\ \citenamefont {Tutis}}]{9}%
  \BibitemOpen
  \bibfield  {author} {\bibinfo {author} {\bibfnamefont {B.}~\bibnamefont
  {Sipos}}, \bibinfo {author} {\bibfnamefont {A.~F.}\ \bibnamefont
  {Kusmartseva}}, \bibinfo {author} {\bibfnamefont {A.}~\bibnamefont {Akrap}},
  \bibinfo {author} {\bibfnamefont {H.}~\bibnamefont {Berger}}, \bibinfo
  {author} {\bibfnamefont {L.}~\bibnamefont {Forro}}, \ and\ \bibinfo {author}
  {\bibfnamefont {E.}~\bibnamefont {Tutis}},\ }\href@noop {} {\bibfield
  {journal} {\bibinfo  {journal} {Nat. Mater.}\ }\textbf {\bibinfo {volume}
  {7}},\ \bibinfo {pages} {960} (\bibinfo {year} {2008})}\BibitemShut {NoStop}%
\bibitem [{\citenamefont {Morosan}\ \emph {et~al.}(2006)\citenamefont
  {Morosan}, \citenamefont {Zandbergen}, \citenamefont {Dennis}, \citenamefont
  {Bos}, \citenamefont {Onose}, \citenamefont {Klimczuk}, \citenamefont
  {Ramirez}, \citenamefont {Ong},\ and\ \citenamefont {Cava}}]{8}%
  \BibitemOpen
  \bibfield  {author} {\bibinfo {author} {\bibfnamefont {E.}~\bibnamefont
  {Morosan}}, \bibinfo {author} {\bibfnamefont {H.~W.}\ \bibnamefont
  {Zandbergen}}, \bibinfo {author} {\bibfnamefont {B.~S.}\ \bibnamefont
  {Dennis}}, \bibinfo {author} {\bibfnamefont {J.~W.~G.}\ \bibnamefont {Bos}},
  \bibinfo {author} {\bibfnamefont {Y.}~\bibnamefont {Onose}}, \bibinfo
  {author} {\bibfnamefont {T.}~\bibnamefont {Klimczuk}}, \bibinfo {author}
  {\bibfnamefont {A.~P.}\ \bibnamefont {Ramirez}}, \bibinfo {author}
  {\bibfnamefont {N.~P.}\ \bibnamefont {Ong}}, \ and\ \bibinfo {author}
  {\bibfnamefont {R.~J.}\ \bibnamefont {Cava}},\ }\href@noop {} {\bibfield
  {journal} {\bibinfo  {journal} {Nat. Phys.}\ }\textbf {\bibinfo {volume}
  {2}},\ \bibinfo {pages} {544} (\bibinfo {year} {2006})}\BibitemShut {NoStop}%
\bibitem [{\citenamefont {Ang}\ \emph {et~al.}(2012)\citenamefont {Ang},
  \citenamefont {Tanaka}, \citenamefont {Ieki}, \citenamefont {Nakayama},
  \citenamefont {Sato}, \citenamefont {Li}, \citenamefont {Lu}, \citenamefont
  {Sun},\ and\ \citenamefont {Takahashi}}]{10}%
  \BibitemOpen
  \bibfield  {author} {\bibinfo {author} {\bibfnamefont {R.}~\bibnamefont
  {Ang}}, \bibinfo {author} {\bibfnamefont {Y.}~\bibnamefont {Tanaka}},
  \bibinfo {author} {\bibfnamefont {E.}~\bibnamefont {Ieki}}, \bibinfo {author}
  {\bibfnamefont {K.}~\bibnamefont {Nakayama}}, \bibinfo {author}
  {\bibfnamefont {T.}~\bibnamefont {Sato}}, \bibinfo {author} {\bibfnamefont
  {L.~J.}\ \bibnamefont {Li}}, \bibinfo {author} {\bibfnamefont {W.~J.}\
  \bibnamefont {Lu}}, \bibinfo {author} {\bibfnamefont {Y.~P.}\ \bibnamefont
  {Sun}}, \ and\ \bibinfo {author} {\bibfnamefont {T.}~\bibnamefont
  {Takahashi}},\ }\href@noop {} {\bibfield  {journal} {\bibinfo  {journal}
  {Phys. Rev. Lett.}\ }\textbf {\bibinfo {volume} {109}},\ \bibinfo {pages}
  {176403} (\bibinfo {year} {2012})}\BibitemShut {NoStop}%
\bibitem [{\citenamefont {Li}\ \emph {et~al.}(2012)\citenamefont {Li},
  \citenamefont {Lu}, \citenamefont {Zhu}, \citenamefont {Ling}, \citenamefont
  {Qu},\ and\ \citenamefont {Sun}}]{11}%
  \BibitemOpen
  \bibfield  {author} {\bibinfo {author} {\bibfnamefont {L.~J.}\ \bibnamefont
  {Li}}, \bibinfo {author} {\bibfnamefont {W.~J.}\ \bibnamefont {Lu}}, \bibinfo
  {author} {\bibfnamefont {X.~D.}\ \bibnamefont {Zhu}}, \bibinfo {author}
  {\bibfnamefont {L.~S.}\ \bibnamefont {Ling}}, \bibinfo {author}
  {\bibfnamefont {Z.}~\bibnamefont {Qu}}, \ and\ \bibinfo {author}
  {\bibfnamefont {Y.~P.}\ \bibnamefont {Sun}},\ }\href@noop {} {\bibfield
  {journal} {\bibinfo  {journal} {Europhys. Lett.}\ }\textbf {\bibinfo {volume}
  {97}},\ \bibinfo {pages} {67005} (\bibinfo {year} {2012})}\BibitemShut
  {NoStop}%
\bibitem [{\citenamefont {Yu}\ \emph {et~al.}(2015)\citenamefont {Yu},
  \citenamefont {Yang}, \citenamefont {Lu}, \citenamefont {Yan}, \citenamefont
  {Cho}, \citenamefont {Ma}, \citenamefont {Niu}, \citenamefont {Kim},
  \citenamefont {Son}, \citenamefont {Feng}, \citenamefont {Li}, \citenamefont
  {Cheong}, \citenamefont {Chen},\ and\ \citenamefont {Zhang}}]{15}%
  \BibitemOpen
  \bibfield  {author} {\bibinfo {author} {\bibfnamefont {Y.}~\bibnamefont
  {Yu}}, \bibinfo {author} {\bibfnamefont {F.}~\bibnamefont {Yang}}, \bibinfo
  {author} {\bibfnamefont {X.~F.}\ \bibnamefont {Lu}}, \bibinfo {author}
  {\bibfnamefont {Y.~J.}\ \bibnamefont {Yan}}, \bibinfo {author} {\bibfnamefont
  {Y.-H.}\ \bibnamefont {Cho}}, \bibinfo {author} {\bibfnamefont
  {L.}~\bibnamefont {Ma}}, \bibinfo {author} {\bibfnamefont {X.}~\bibnamefont
  {Niu}}, \bibinfo {author} {\bibfnamefont {S.}~\bibnamefont {Kim}}, \bibinfo
  {author} {\bibfnamefont {Y.-W.}\ \bibnamefont {Son}}, \bibinfo {author}
  {\bibfnamefont {D.}~\bibnamefont {Feng}}, \bibinfo {author} {\bibfnamefont
  {S.}~\bibnamefont {Li}}, \bibinfo {author} {\bibfnamefont {S.-W.}\
  \bibnamefont {Cheong}}, \bibinfo {author} {\bibfnamefont {X.~H.}\
  \bibnamefont {Chen}}, \ and\ \bibinfo {author} {\bibfnamefont
  {Y.}~\bibnamefont {Zhang}},\ }\href@noop {} {\bibfield  {journal} {\bibinfo
  {journal} {Nat. Nanotech.}\ }\textbf {\bibinfo {volume} {10}},\ \bibinfo
  {pages} {270} (\bibinfo {year} {2015})}\BibitemShut {NoStop}%
\bibitem [{\citenamefont {Ang}\ \emph {et~al.}(2013)\citenamefont {Ang},
  \citenamefont {Miyata}, \citenamefont {Ieki}, \citenamefont {Nakayama},
  \citenamefont {Sato}, \citenamefont {Liu}, \citenamefont {Lu}, \citenamefont
  {Sun},\ and\ \citenamefont {Takahashi}}]{12}%
  \BibitemOpen
  \bibfield  {author} {\bibinfo {author} {\bibfnamefont {R.}~\bibnamefont
  {Ang}}, \bibinfo {author} {\bibfnamefont {Y.}~\bibnamefont {Miyata}},
  \bibinfo {author} {\bibfnamefont {E.}~\bibnamefont {Ieki}}, \bibinfo {author}
  {\bibfnamefont {K.}~\bibnamefont {Nakayama}}, \bibinfo {author}
  {\bibfnamefont {T.}~\bibnamefont {Sato}}, \bibinfo {author} {\bibfnamefont
  {Y.}~\bibnamefont {Liu}}, \bibinfo {author} {\bibfnamefont {W.~J.}\
  \bibnamefont {Lu}}, \bibinfo {author} {\bibfnamefont {Y.~P.}\ \bibnamefont
  {Sun}}, \ and\ \bibinfo {author} {\bibfnamefont {T.}~\bibnamefont
  {Takahashi}},\ }\href@noop {} {\bibfield  {journal} {\bibinfo  {journal}
  {Phys. Rev. B}\ }\textbf {\bibinfo {volume} {88}},\ \bibinfo {pages} {115145}
  (\bibinfo {year} {2013})}\BibitemShut {NoStop}%
\bibitem [{\citenamefont {Liu}\ \emph {et~al.}(2013)\citenamefont {Liu},
  \citenamefont {Ang}, \citenamefont {Lu}, \citenamefont {Song}, \citenamefont
  {Li},\ and\ \citenamefont {Sun}}]{13}%
  \BibitemOpen
  \bibfield  {author} {\bibinfo {author} {\bibfnamefont {Y.}~\bibnamefont
  {Liu}}, \bibinfo {author} {\bibfnamefont {R.}~\bibnamefont {Ang}}, \bibinfo
  {author} {\bibfnamefont {W.~J.}\ \bibnamefont {Lu}}, \bibinfo {author}
  {\bibfnamefont {W.~H.}\ \bibnamefont {Song}}, \bibinfo {author}
  {\bibfnamefont {L.~J.}\ \bibnamefont {Li}}, \ and\ \bibinfo {author}
  {\bibfnamefont {Y.~P.}\ \bibnamefont {Sun}},\ }\href@noop {} {\bibfield
  {journal} {\bibinfo  {journal} {Appl. Phys. Lett.}\ }\textbf {\bibinfo
  {volume} {102}},\ \bibinfo {pages} {192602} (\bibinfo {year}
  {2013})}\BibitemShut {NoStop}%
\bibitem [{\citenamefont {Ang}\ \emph {et~al.}(2015)\citenamefont {Ang},
  \citenamefont {Wang}, \citenamefont {Chen}, \citenamefont {Tang},
  \citenamefont {Liu}, \citenamefont {Liu}, \citenamefont {Lu}, \citenamefont
  {Sun}, \citenamefont {Mori},\ and\ \citenamefont {Ikuhara}}]{14}%
  \BibitemOpen
  \bibfield  {author} {\bibinfo {author} {\bibfnamefont {R.}~\bibnamefont
  {Ang}}, \bibinfo {author} {\bibfnamefont {Z.~C.}\ \bibnamefont {Wang}},
  \bibinfo {author} {\bibfnamefont {C.~L.}\ \bibnamefont {Chen}}, \bibinfo
  {author} {\bibfnamefont {J.}~\bibnamefont {Tang}}, \bibinfo {author}
  {\bibfnamefont {N.}~\bibnamefont {Liu}}, \bibinfo {author} {\bibfnamefont
  {Y.}~\bibnamefont {Liu}}, \bibinfo {author} {\bibfnamefont {W.~J.}\
  \bibnamefont {Lu}}, \bibinfo {author} {\bibfnamefont {Y.~P.}\ \bibnamefont
  {Sun}}, \bibinfo {author} {\bibfnamefont {T.}~\bibnamefont {Mori}}, \ and\
  \bibinfo {author} {\bibfnamefont {Y.}~\bibnamefont {Ikuhara}},\ }\href@noop
  {} {\bibfield  {journal} {\bibinfo  {journal} {Nat. Commun.}\ }\textbf
  {\bibinfo {volume} {6}},\ \bibinfo {pages} {6091} (\bibinfo {year}
  {2015})}\BibitemShut {NoStop}%
\bibitem [{\citenamefont {Wilson}\ \emph {et~al.}(1975)\citenamefont {Wilson},
  \citenamefont {DiSalvo},\ and\ \citenamefont {Mahajan}}]{1}%
  \BibitemOpen
  \bibfield  {author} {\bibinfo {author} {\bibfnamefont {J.~A.}\ \bibnamefont
  {Wilson}}, \bibinfo {author} {\bibfnamefont {F.~J.}\ \bibnamefont {DiSalvo}},
  \ and\ \bibinfo {author} {\bibfnamefont {S.}~\bibnamefont {Mahajan}},\
  }\href@noop {} {\bibfield  {journal} {\bibinfo  {journal} {Adv. Phys.}\
  }\textbf {\bibinfo {volume} {24}},\ \bibinfo {pages} {117} (\bibinfo {year}
  {1975})}\BibitemShut {NoStop}%
\bibitem [{\citenamefont {Scruby}\ \emph {et~al.}(1975)\citenamefont {Scruby},
  \citenamefont {Williams},\ and\ \citenamefont {Parry}}]{2}%
  \BibitemOpen
  \bibfield  {author} {\bibinfo {author} {\bibfnamefont {C.~B.}\ \bibnamefont
  {Scruby}}, \bibinfo {author} {\bibfnamefont {P.~M.}\ \bibnamefont
  {Williams}}, \ and\ \bibinfo {author} {\bibfnamefont {G.~S.}\ \bibnamefont
  {Parry}},\ }\href@noop {} {\bibfield  {journal} {\bibinfo  {journal} {Philos.
  Mag.}\ }\textbf {\bibinfo {volume} {31}},\ \bibinfo {pages} {255} (\bibinfo
  {year} {1975})}\BibitemShut {NoStop}%
\bibitem [{\citenamefont {Fazekas}\ and\ \citenamefont {Tosatti}(1979)}]{3}%
  \BibitemOpen
  \bibfield  {author} {\bibinfo {author} {\bibfnamefont {P.}~\bibnamefont
  {Fazekas}}\ and\ \bibinfo {author} {\bibfnamefont {E.}~\bibnamefont
  {Tosatti}},\ }\href@noop {} {\bibfield  {journal} {\bibinfo  {journal}
  {Philos. Mag. B}\ }\textbf {\bibinfo {volume} {39}},\ \bibinfo {pages} {229}
  (\bibinfo {year} {1979})}\BibitemShut {NoStop}%
\bibitem [{\citenamefont {Fazekas}\ and\ \citenamefont {Tosatti}(1980)}]{4}%
  \BibitemOpen
  \bibfield  {author} {\bibinfo {author} {\bibfnamefont {P.}~\bibnamefont
  {Fazekas}}\ and\ \bibinfo {author} {\bibfnamefont {E.}~\bibnamefont
  {Tosatti}},\ }\href@noop {} {\bibfield  {journal} {\bibinfo  {journal}
  {Physica B}\ }\textbf {\bibinfo {volume} {99}},\ \bibinfo {pages} {183}
  (\bibinfo {year} {1980})}\BibitemShut {NoStop}%
\bibitem [{\citenamefont {Dardel}\ \emph
  {et~al.}(1992{\natexlab{a}})\citenamefont {Dardel}, \citenamefont {Grioni},
  \citenamefont {Malterre}, \citenamefont {Weibel}, \citenamefont {Baer},\ and\
  \citenamefont {Levy}}]{5}%
  \BibitemOpen
  \bibfield  {author} {\bibinfo {author} {\bibfnamefont {B.}~\bibnamefont
  {Dardel}}, \bibinfo {author} {\bibfnamefont {M.}~\bibnamefont {Grioni}},
  \bibinfo {author} {\bibfnamefont {D.}~\bibnamefont {Malterre}}, \bibinfo
  {author} {\bibfnamefont {P.}~\bibnamefont {Weibel}}, \bibinfo {author}
  {\bibfnamefont {Y.}~\bibnamefont {Baer}}, \ and\ \bibinfo {author}
  {\bibfnamefont {F.}~\bibnamefont {Levy}},\ }\href@noop {} {\bibfield
  {journal} {\bibinfo  {journal} {Phys. Rev. B}\ }\textbf {\bibinfo {volume}
  {45}},\ \bibinfo {pages} {1462} (\bibinfo {year}
  {1992}{\natexlab{a}})}\BibitemShut {NoStop}%
\bibitem [{\citenamefont {Dardel}\ \emph
  {et~al.}(1992{\natexlab{b}})\citenamefont {Dardel}, \citenamefont {Grioni},
  \citenamefont {Malterre}, \citenamefont {Weibel}, \citenamefont {Baer},\ and\
  \citenamefont {Levy}}]{6}%
  \BibitemOpen
  \bibfield  {author} {\bibinfo {author} {\bibfnamefont {B.}~\bibnamefont
  {Dardel}}, \bibinfo {author} {\bibfnamefont {M.}~\bibnamefont {Grioni}},
  \bibinfo {author} {\bibfnamefont {D.}~\bibnamefont {Malterre}}, \bibinfo
  {author} {\bibfnamefont {P.}~\bibnamefont {Weibel}}, \bibinfo {author}
  {\bibfnamefont {Y.}~\bibnamefont {Baer}}, \ and\ \bibinfo {author}
  {\bibfnamefont {F.}~\bibnamefont {Levy}},\ }\href@noop {} {\bibfield
  {journal} {\bibinfo  {journal} {Phys. Rev. B}\ }\textbf {\bibinfo {volume}
  {46}},\ \bibinfo {pages} {7407} (\bibinfo {year}
  {1992}{\natexlab{b}})}\BibitemShut {NoStop}%
\bibitem [{\citenamefont {Kim}\ \emph {et~al.}(1994)\citenamefont {Kim},
  \citenamefont {Yamaguchi}, \citenamefont {Hasegawa},\ and\ \citenamefont
  {Kitazawa}}]{7}%
  \BibitemOpen
  \bibfield  {author} {\bibinfo {author} {\bibfnamefont {J.-J.}\ \bibnamefont
  {Kim}}, \bibinfo {author} {\bibfnamefont {W.}~\bibnamefont {Yamaguchi}},
  \bibinfo {author} {\bibfnamefont {T.}~\bibnamefont {Hasegawa}}, \ and\
  \bibinfo {author} {\bibfnamefont {K.}~\bibnamefont {Kitazawa}},\ }\href@noop
  {} {\bibfield  {journal} {\bibinfo  {journal} {Phys. Rev. Lett.}\ }\textbf
  {\bibinfo {volume} {73}},\ \bibinfo {pages} {2103} (\bibinfo {year}
  {1994})}\BibitemShut {NoStop}%
\bibitem [{\citenamefont {Wu}\ and\ \citenamefont {Lieber}(1989)}]{50}%
  \BibitemOpen
  \bibfield  {author} {\bibinfo {author} {\bibfnamefont {X.~L.}\ \bibnamefont
  {Wu}}\ and\ \bibinfo {author} {\bibfnamefont {C.~M.}\ \bibnamefont
  {Lieber}},\ }\href@noop {} {\bibfield  {journal} {\bibinfo  {journal}
  {Science}\ }\textbf {\bibinfo {volume} {243}},\ \bibinfo {pages} {1703}
  (\bibinfo {year} {1989})}\BibitemShut {NoStop}%
\bibitem [{\citenamefont {Lahoud}\ \emph {et~al.}(2014)\citenamefont {Lahoud},
  \citenamefont {Meetei}, \citenamefont {Chaska}, \citenamefont {Kanigel},\
  and\ \citenamefont {Trivedi}}]{20}%
  \BibitemOpen
  \bibfield  {author} {\bibinfo {author} {\bibfnamefont {E.}~\bibnamefont
  {Lahoud}}, \bibinfo {author} {\bibfnamefont {O.~N.}\ \bibnamefont {Meetei}},
  \bibinfo {author} {\bibfnamefont {K.~B.}\ \bibnamefont {Chaska}}, \bibinfo
  {author} {\bibfnamefont {A.}~\bibnamefont {Kanigel}}, \ and\ \bibinfo
  {author} {\bibfnamefont {N.}~\bibnamefont {Trivedi}},\ }\href@noop {}
  {\bibfield  {journal} {\bibinfo  {journal} {Phys. Rev. Lett.}\ }\textbf
  {\bibinfo {volume} {112}},\ \bibinfo {pages} {206402} (\bibinfo {year}
  {2014})}\BibitemShut {NoStop}%
\bibitem [{\citenamefont {Aryanpour}\ \emph {et~al.}(2006)\citenamefont
  {Aryanpour}, \citenamefont {Pickett},\ and\ \citenamefont {Scalettar}}]{30}%
  \BibitemOpen
  \bibfield  {author} {\bibinfo {author} {\bibfnamefont {K.}~\bibnamefont
  {Aryanpour}}, \bibinfo {author} {\bibfnamefont {W.~E.}\ \bibnamefont
  {Pickett}}, \ and\ \bibinfo {author} {\bibfnamefont {R.~T.}\ \bibnamefont
  {Scalettar}},\ }\href@noop {} {\bibfield  {journal} {\bibinfo  {journal}
  {Phys. Rev. B}\ }\textbf {\bibinfo {volume} {74}},\ \bibinfo {pages} {085117}
  (\bibinfo {year} {2006})}\BibitemShut {NoStop}%
\bibitem [{\citenamefont {Perfetti}\ \emph {et~al.}(2006)\citenamefont
  {Perfetti}, \citenamefont {Loukakos}, \citenamefont {Lisowski}, \citenamefont
  {Bovensiepen}, \citenamefont {Berger}, \citenamefont {Biermann},
  \citenamefont {Cornaglia}, \citenamefont {Georges},\ and\ \citenamefont
  {Wolf}}]{31}%
  \BibitemOpen
  \bibfield  {author} {\bibinfo {author} {\bibfnamefont {L.}~\bibnamefont
  {Perfetti}}, \bibinfo {author} {\bibfnamefont {P.~A.}\ \bibnamefont
  {Loukakos}}, \bibinfo {author} {\bibfnamefont {M.}~\bibnamefont {Lisowski}},
  \bibinfo {author} {\bibfnamefont {U.}~\bibnamefont {Bovensiepen}}, \bibinfo
  {author} {\bibfnamefont {H.}~\bibnamefont {Berger}}, \bibinfo {author}
  {\bibfnamefont {S.}~\bibnamefont {Biermann}}, \bibinfo {author}
  {\bibfnamefont {P.~S.}\ \bibnamefont {Cornaglia}}, \bibinfo {author}
  {\bibfnamefont {A.}~\bibnamefont {Georges}}, \ and\ \bibinfo {author}
  {\bibfnamefont {M.}~\bibnamefont {Wolf}},\ }\href@noop {} {\bibfield
  {journal} {\bibinfo  {journal} {Phys. Rev. Lett.}\ }\textbf {\bibinfo
  {volume} {97}},\ \bibinfo {pages} {067402} (\bibinfo {year}
  {2006})}\BibitemShut {NoStop}%
\bibitem [{\citenamefont {Mutka}\ \emph {et~al.}(1981)\citenamefont {Mutka},
  \citenamefont {Zuppiroli}, \citenamefont {Molinie},\ and\ \citenamefont
  {Bourgoin}}]{16}%
  \BibitemOpen
  \bibfield  {author} {\bibinfo {author} {\bibfnamefont {H.}~\bibnamefont
  {Mutka}}, \bibinfo {author} {\bibfnamefont {L.}~\bibnamefont {Zuppiroli}},
  \bibinfo {author} {\bibfnamefont {P.}~\bibnamefont {Molinie}}, \ and\
  \bibinfo {author} {\bibfnamefont {J.~C.}\ \bibnamefont {Bourgoin}},\
  }\href@noop {} {\bibfield  {journal} {\bibinfo  {journal} {Phys. Rev. B}\
  }\textbf {\bibinfo {volume} {23}},\ \bibinfo {pages} {5030} (\bibinfo {year}
  {1981})}\BibitemShut {NoStop}%
\bibitem [{\citenamefont {Bovet}\ \emph {et~al.}(2003)\citenamefont {Bovet},
  \citenamefont {vanSmaalen}, \citenamefont {Berger}, \citenamefont {Gaal},
  \citenamefont {Forro}, \citenamefont {Schlapbach},\ and\ \citenamefont
  {Aebi}}]{17}%
  \BibitemOpen
  \bibfield  {author} {\bibinfo {author} {\bibfnamefont {M.}~\bibnamefont
  {Bovet}}, \bibinfo {author} {\bibfnamefont {S.}~\bibnamefont {vanSmaalen}},
  \bibinfo {author} {\bibfnamefont {H.}~\bibnamefont {Berger}}, \bibinfo
  {author} {\bibfnamefont {R.}~\bibnamefont {Gaal}}, \bibinfo {author}
  {\bibfnamefont {L.}~\bibnamefont {Forro}}, \bibinfo {author} {\bibfnamefont
  {L.}~\bibnamefont {Schlapbach}}, \ and\ \bibinfo {author} {\bibfnamefont
  {P.}~\bibnamefont {Aebi}},\ }\href@noop {} {\bibfield  {journal} {\bibinfo
  {journal} {Phys. Rev. B}\ }\textbf {\bibinfo {volume} {67}},\ \bibinfo
  {pages} {125105} (\bibinfo {year} {2003})}\BibitemShut {NoStop}%
\bibitem [{\citenamefont {Rossnagel}\ and\ \citenamefont {Smith}(2006)}]{18}%
  \BibitemOpen
  \bibfield  {author} {\bibinfo {author} {\bibfnamefont {K.}~\bibnamefont
  {Rossnagel}}\ and\ \bibinfo {author} {\bibfnamefont {N.~V.}\ \bibnamefont
  {Smith}},\ }\href@noop {} {\bibfield  {journal} {\bibinfo  {journal} {Phys.
  Rev. B}\ }\textbf {\bibinfo {volume} {73}},\ \bibinfo {pages} {073106}
  (\bibinfo {year} {2006})}\BibitemShut {NoStop}%
\bibitem [{\citenamefont {Xu}\ \emph {et~al.}(2010)\citenamefont {Xu},
  \citenamefont {Piatek}, \citenamefont {Lin}, \citenamefont {Sipos},
  \citenamefont {Berger}, \citenamefont {Forro}, \citenamefont {Ronnow},\ and\
  \citenamefont {Grioni}}]{19}%
  \BibitemOpen
  \bibfield  {author} {\bibinfo {author} {\bibfnamefont {P.}~\bibnamefont
  {Xu}}, \bibinfo {author} {\bibfnamefont {J.~O.}\ \bibnamefont {Piatek}},
  \bibinfo {author} {\bibfnamefont {P.-H.}\ \bibnamefont {Lin}}, \bibinfo
  {author} {\bibfnamefont {B.}~\bibnamefont {Sipos}}, \bibinfo {author}
  {\bibfnamefont {H.}~\bibnamefont {Berger}}, \bibinfo {author} {\bibfnamefont
  {L.}~\bibnamefont {Forro}}, \bibinfo {author} {\bibfnamefont {H.~M.}\
  \bibnamefont {Ronnow}}, \ and\ \bibinfo {author} {\bibfnamefont
  {M.}~\bibnamefont {Grioni}},\ }\href@noop {} {\bibfield  {journal} {\bibinfo
  {journal} {Phys. Rev. B}\ }\textbf {\bibinfo {volume} {81}},\ \bibinfo
  {pages} {172503} (\bibinfo {year} {2010})}\BibitemShut {NoStop}%
\bibitem [{\citenamefont {Cho}\ \emph {et~al.}(2015)\citenamefont {Cho},
  \citenamefont {Cho}, \citenamefont {Cheong}, \citenamefont {Kim},\ and\
  \citenamefont {Yeom}}]{21}%
  \BibitemOpen
  \bibfield  {author} {\bibinfo {author} {\bibfnamefont {D.}~\bibnamefont
  {Cho}}, \bibinfo {author} {\bibfnamefont {Y.-H.}\ \bibnamefont {Cho}},
  \bibinfo {author} {\bibfnamefont {S.-W.}\ \bibnamefont {Cheong}}, \bibinfo
  {author} {\bibfnamefont {K.-S.}\ \bibnamefont {Kim}}, \ and\ \bibinfo
  {author} {\bibfnamefont {H.~W.}\ \bibnamefont {Yeom}},\ }\href@noop {}
  {\bibfield  {journal} {\bibinfo  {journal} {Phys. Rev. B}\ }\textbf {\bibinfo
  {volume} {92}},\ \bibinfo {pages} {085132} (\bibinfo {year}
  {2015})}\BibitemShut {NoStop}%
\bibitem [{\citenamefont {Ritschel}\ \emph {et~al.}(2015)\citenamefont
  {Ritschel}, \citenamefont {Trinckauf}, \citenamefont {Koepernik},
  \citenamefont {Buchner}, \citenamefont {v.~Zimmermann}, \citenamefont
  {Berger}, \citenamefont {Joe}, \citenamefont {Abbamonte},\ and\ \citenamefont
  {Geck}}]{22}%
  \BibitemOpen
  \bibfield  {author} {\bibinfo {author} {\bibfnamefont {T.}~\bibnamefont
  {Ritschel}}, \bibinfo {author} {\bibfnamefont {J.}~\bibnamefont {Trinckauf}},
  \bibinfo {author} {\bibfnamefont {K.}~\bibnamefont {Koepernik}}, \bibinfo
  {author} {\bibfnamefont {B.}~\bibnamefont {Buchner}}, \bibinfo {author}
  {\bibfnamefont {M.}~\bibnamefont {v.~Zimmermann}}, \bibinfo {author}
  {\bibfnamefont {H.}~\bibnamefont {Berger}}, \bibinfo {author} {\bibfnamefont
  {Y.~I.}\ \bibnamefont {Joe}}, \bibinfo {author} {\bibfnamefont
  {P.}~\bibnamefont {Abbamonte}}, \ and\ \bibinfo {author} {\bibfnamefont
  {J.}~\bibnamefont {Geck}},\ }\href@noop {} {\bibfield  {journal} {\bibinfo
  {journal} {Nat. Phys.}\ }\textbf {\bibinfo {volume} {11}},\ \bibinfo {pages}
  {328} (\bibinfo {year} {2015})}\BibitemShut {NoStop}%
\bibitem [{\citenamefont {Ye}\ \emph {et~al.}(2013)\citenamefont {Ye},
  \citenamefont {Cai}, \citenamefont {Yu}, \citenamefont {Zhou}, \citenamefont
  {Ruan}, \citenamefont {Liu}, \citenamefont {Jin},\ and\ \citenamefont
  {Wang}}]{34}%
  \BibitemOpen
  \bibfield  {author} {\bibinfo {author} {\bibfnamefont {C.}~\bibnamefont
  {Ye}}, \bibinfo {author} {\bibfnamefont {P.}~\bibnamefont {Cai}}, \bibinfo
  {author} {\bibfnamefont {R.}~\bibnamefont {Yu}}, \bibinfo {author}
  {\bibfnamefont {X.}~\bibnamefont {Zhou}}, \bibinfo {author} {\bibfnamefont
  {W.}~\bibnamefont {Ruan}}, \bibinfo {author} {\bibfnamefont {Q.}~\bibnamefont
  {Liu}}, \bibinfo {author} {\bibfnamefont {C.}~\bibnamefont {Jin}}, \ and\
  \bibinfo {author} {\bibfnamefont {Y.}~\bibnamefont {Wang}},\ }\href@noop {}
  {\bibfield  {journal} {\bibinfo  {journal} {Nat. Commun.}\ }\textbf {\bibinfo
  {volume} {4}},\ \bibinfo {pages} {1365} (\bibinfo {year} {2013})}\BibitemShut
  {NoStop}%
\bibitem [{\citenamefont {Kresse}\ and\ \citenamefont {Hafner}(1993)}]{35}%
  \BibitemOpen
  \bibfield  {author} {\bibinfo {author} {\bibfnamefont {G.}~\bibnamefont
  {Kresse}}\ and\ \bibinfo {author} {\bibfnamefont {J.}~\bibnamefont
  {Hafner}},\ }\href@noop {} {\bibfield  {journal} {\bibinfo  {journal} {Phys.
  Rev. B}\ }\textbf {\bibinfo {volume} {47}},\ \bibinfo {pages} {558} (\bibinfo
  {year} {1993})}\BibitemShut {NoStop}%
\bibitem [{\citenamefont {Kresse}\ and\ \citenamefont {Hafner}(1994)}]{36}%
  \BibitemOpen
  \bibfield  {author} {\bibinfo {author} {\bibfnamefont {G.}~\bibnamefont
  {Kresse}}\ and\ \bibinfo {author} {\bibfnamefont {J.}~\bibnamefont
  {Hafner}},\ }\href@noop {} {\bibfield  {journal} {\bibinfo  {journal} {Phys.
  Rev. B}\ }\textbf {\bibinfo {volume} {49}},\ \bibinfo {pages} {14251}
  (\bibinfo {year} {1994})}\BibitemShut {NoStop}%
\bibitem [{\citenamefont {Kresse}\ and\ \citenamefont
  {Furthmuller}(1996{\natexlab{a}})}]{37}%
  \BibitemOpen
  \bibfield  {author} {\bibinfo {author} {\bibfnamefont {G.}~\bibnamefont
  {Kresse}}\ and\ \bibinfo {author} {\bibfnamefont {J.}~\bibnamefont
  {Furthmuller}},\ }\href@noop {} {\bibfield  {journal} {\bibinfo  {journal}
  {Comput. Mater. Sci.}\ }\textbf {\bibinfo {volume} {6}},\ \bibinfo {pages}
  {15} (\bibinfo {year} {1996}{\natexlab{a}})}\BibitemShut {NoStop}%
\bibitem [{\citenamefont {Kresse}\ and\ \citenamefont
  {Furthmuller}(1996{\natexlab{b}})}]{38}%
  \BibitemOpen
  \bibfield  {author} {\bibinfo {author} {\bibfnamefont {G.}~\bibnamefont
  {Kresse}}\ and\ \bibinfo {author} {\bibfnamefont {J.}~\bibnamefont
  {Furthmuller}},\ }\href@noop {} {\bibfield  {journal} {\bibinfo  {journal}
  {Phys. Rev. B}\ }\textbf {\bibinfo {volume} {54}},\ \bibinfo {pages} {11169}
  (\bibinfo {year} {1996}{\natexlab{b}})}\BibitemShut {NoStop}%
\bibitem [{\citenamefont {Perdew}\ \emph {et~al.}(1996)\citenamefont {Perdew},
  \citenamefont {Burke},\ and\ \citenamefont {Ernzerhof}}]{39}%
  \BibitemOpen
  \bibfield  {author} {\bibinfo {author} {\bibfnamefont {J.~P.}\ \bibnamefont
  {Perdew}}, \bibinfo {author} {\bibfnamefont {K.}~\bibnamefont {Burke}}, \
  and\ \bibinfo {author} {\bibfnamefont {M.}~\bibnamefont {Ernzerhof}},\
  }\href@noop {} {\bibfield  {journal} {\bibinfo  {journal} {Phys. Rev. Lett.}\
  }\textbf {\bibinfo {volume} {77}},\ \bibinfo {pages} {3865} (\bibinfo {year}
  {1996})}\BibitemShut {NoStop}%
\bibitem [{\citenamefont {Kresse}\ and\ \citenamefont {Joubert}(1999)}]{40}%
  \BibitemOpen
  \bibfield  {author} {\bibinfo {author} {\bibfnamefont {G.}~\bibnamefont
  {Kresse}}\ and\ \bibinfo {author} {\bibfnamefont {D.}~\bibnamefont
  {Joubert}},\ }\href@noop {} {\bibfield  {journal} {\bibinfo  {journal} {Phys.
  Rev. B}\ }\textbf {\bibinfo {volume} {59}},\ \bibinfo {pages} {1758}
  (\bibinfo {year} {1999})}\BibitemShut {NoStop}%
\bibitem [{\citenamefont {Marzari}\ and\ \citenamefont
  {Vanderbilt}(1997)}]{27}%
  \BibitemOpen
  \bibfield  {author} {\bibinfo {author} {\bibfnamefont {N.}~\bibnamefont
  {Marzari}}\ and\ \bibinfo {author} {\bibfnamefont {D.}~\bibnamefont
  {Vanderbilt}},\ }\href@noop {} {\bibfield  {journal} {\bibinfo  {journal}
  {Phys. Rev. B}\ }\textbf {\bibinfo {volume} {56}},\ \bibinfo {pages} {12847}
  (\bibinfo {year} {1997})}\BibitemShut {NoStop}%
\bibitem [{\citenamefont {Souza}\ \emph {et~al.}(2001)\citenamefont {Souza},
  \citenamefont {Marzari},\ and\ \citenamefont {Vanderbilt}}]{28}%
  \BibitemOpen
  \bibfield  {author} {\bibinfo {author} {\bibfnamefont {I.}~\bibnamefont
  {Souza}}, \bibinfo {author} {\bibfnamefont {N.}~\bibnamefont {Marzari}}, \
  and\ \bibinfo {author} {\bibfnamefont {D.}~\bibnamefont {Vanderbilt}},\
  }\href@noop {} {\bibfield  {journal} {\bibinfo  {journal} {Phys. Rev. B}\
  }\textbf {\bibinfo {volume} {65}},\ \bibinfo {pages} {035109} (\bibinfo
  {year} {2001})}\BibitemShut {NoStop}%
\bibitem [{\citenamefont {Mostofi}\ \emph {et~al.}(2008)\citenamefont
  {Mostofi}, \citenamefont {Yates}, \citenamefont {Lee}, \citenamefont {Souza},
  \citenamefont {Vanderbilt},\ and\ \citenamefont {Marzari}}]{29}%
  \BibitemOpen
  \bibfield  {author} {\bibinfo {author} {\bibfnamefont {A.~A.}\ \bibnamefont
  {Mostofi}}, \bibinfo {author} {\bibfnamefont {J.~R.}\ \bibnamefont {Yates}},
  \bibinfo {author} {\bibfnamefont {Y.-S.}\ \bibnamefont {Lee}}, \bibinfo
  {author} {\bibfnamefont {I.}~\bibnamefont {Souza}}, \bibinfo {author}
  {\bibfnamefont {D.}~\bibnamefont {Vanderbilt}}, \ and\ \bibinfo {author}
  {\bibfnamefont {N.}~\bibnamefont {Marzari}},\ }\href@noop {} {\bibfield
  {journal} {\bibinfo  {journal} {Comput. Phys. Commun.}\ }\textbf {\bibinfo
  {volume} {178}},\ \bibinfo {pages} {685} (\bibinfo {year}
  {2008})}\BibitemShut {NoStop}%
\bibitem [{\citenamefont {Liechtenstein}\ \emph {et~al.}(1995)\citenamefont
  {Liechtenstein}, \citenamefont {Anisimov},\ and\ \citenamefont
  {Zaanen}}]{41}%
  \BibitemOpen
  \bibfield  {author} {\bibinfo {author} {\bibfnamefont {A.~I.}\ \bibnamefont
  {Liechtenstein}}, \bibinfo {author} {\bibfnamefont {V.~I.}\ \bibnamefont
  {Anisimov}}, \ and\ \bibinfo {author} {\bibfnamefont {J.}~\bibnamefont
  {Zaanen}},\ }\href@noop {} {\bibfield  {journal} {\bibinfo  {journal} {Phys.
  Rev. B}\ }\textbf {\bibinfo {volume} {52}},\ \bibinfo {pages} {R5467}
  (\bibinfo {year} {1995})}\BibitemShut {NoStop}%
\bibitem [{\citenamefont {Dudarev}\ \emph {et~al.}(1998)\citenamefont
  {Dudarev}, \citenamefont {Botton}, \citenamefont {Savrasov}, \citenamefont
  {Humphreys},\ and\ \citenamefont {Sutton}}]{42}%
  \BibitemOpen
  \bibfield  {author} {\bibinfo {author} {\bibfnamefont {S.~L.}\ \bibnamefont
  {Dudarev}}, \bibinfo {author} {\bibfnamefont {G.~A.}\ \bibnamefont {Botton}},
  \bibinfo {author} {\bibfnamefont {S.~Y.}\ \bibnamefont {Savrasov}}, \bibinfo
  {author} {\bibfnamefont {C.~J.}\ \bibnamefont {Humphreys}}, \ and\ \bibinfo
  {author} {\bibfnamefont {A.~P.}\ \bibnamefont {Sutton}},\ }\href@noop {}
  {\bibfield  {journal} {\bibinfo  {journal} {Phys. Rev. B}\ }\textbf {\bibinfo
  {volume} {57}},\ \bibinfo {pages} {1505} (\bibinfo {year}
  {1998})}\BibitemShut {NoStop}%
\bibitem [{\citenamefont {Jellinek}\ and\ \citenamefont {Less}(1962)}]{43}%
  \BibitemOpen
  \bibfield  {author} {\bibinfo {author} {\bibfnamefont {F.~J.~F.}\
  \bibnamefont {Jellinek}}\ and\ \bibinfo {author} {\bibfnamefont
  {J.}~\bibnamefont {Less}},\ }\href@noop {} {\bibfield  {journal} {\bibinfo
  {journal} {J. Less-Common Met.}\ }\textbf {\bibinfo {volume} {4}},\ \bibinfo
  {pages} {9} (\bibinfo {year} {1962})}\BibitemShut {NoStop}%
\bibitem [{\citenamefont {Myron}\ and\ \citenamefont {Freeman}(1975)}]{44}%
  \BibitemOpen
  \bibfield  {author} {\bibinfo {author} {\bibfnamefont {H.~W.}\ \bibnamefont
  {Myron}}\ and\ \bibinfo {author} {\bibfnamefont {A.~J.}\ \bibnamefont
  {Freeman}},\ }\href@noop {} {\bibfield  {journal} {\bibinfo  {journal} {Phys.
  Rev. B}\ }\textbf {\bibinfo {volume} {11}},\ \bibinfo {pages} {2735}
  (\bibinfo {year} {1975})}\BibitemShut {NoStop}%
\bibitem [{\citenamefont {Woolley}\ and\ \citenamefont {Wexler}(1977)}]{45}%
  \BibitemOpen
  \bibfield  {author} {\bibinfo {author} {\bibfnamefont {A.~M.}\ \bibnamefont
  {Woolley}}\ and\ \bibinfo {author} {\bibfnamefont {G.}~\bibnamefont
  {Wexler}},\ }\href@noop {} {\bibfield  {journal} {\bibinfo  {journal} {J.
  Phys. C: Solid State Phys.}\ }\textbf {\bibinfo {volume} {10}},\ \bibinfo
  {pages} {2601} (\bibinfo {year} {1977})}\BibitemShut {NoStop}%
\bibitem [{\citenamefont {Darancet}\ \emph {et~al.}(2014)\citenamefont
  {Darancet}, \citenamefont {Millis},\ and\ \citenamefont {Marianetti}}]{23}%
  \BibitemOpen
  \bibfield  {author} {\bibinfo {author} {\bibfnamefont {P.}~\bibnamefont
  {Darancet}}, \bibinfo {author} {\bibfnamefont {A.~J.}\ \bibnamefont
  {Millis}}, \ and\ \bibinfo {author} {\bibfnamefont {C.~A.}\ \bibnamefont
  {Marianetti}},\ }\href@noop {} {\bibfield  {journal} {\bibinfo  {journal}
  {Phys. Rev. B}\ }\textbf {\bibinfo {volume} {90}},\ \bibinfo {pages} {045134}
  (\bibinfo {year} {2014})}\BibitemShut {NoStop}%
\bibitem [{\citenamefont {Smith}\ \emph {et~al.}(1985)\citenamefont {Smith},
  \citenamefont {Kevan},\ and\ \citenamefont {DiSalvo}}]{24}%
  \BibitemOpen
  \bibfield  {author} {\bibinfo {author} {\bibfnamefont {N.~V.}\ \bibnamefont
  {Smith}}, \bibinfo {author} {\bibfnamefont {S.~D.}\ \bibnamefont {Kevan}}, \
  and\ \bibinfo {author} {\bibfnamefont {F.~J.}\ \bibnamefont {DiSalvo}},\
  }\href@noop {} {\bibfield  {journal} {\bibinfo  {journal} {J. Phys. C: Solid
  State Phys.}\ }\textbf {\bibinfo {volume} {18}},\ \bibinfo {pages} {3175}
  (\bibinfo {year} {1985})}\BibitemShut {NoStop}%
\bibitem [{\citenamefont {Zwick}\ \emph {et~al.}(1998)\citenamefont {Zwick},
  \citenamefont {Berger}, \citenamefont {Vobornik}, \citenamefont
  {Margaritondo}, \citenamefont {Forro}, \citenamefont {Beeli}, \citenamefont
  {Onellion}, \citenamefont {Panaccione}, \citenamefont {Taleb-Ibrahimi},\ and\
  \citenamefont {Grioni}}]{25}%
  \BibitemOpen
  \bibfield  {author} {\bibinfo {author} {\bibfnamefont {F.}~\bibnamefont
  {Zwick}}, \bibinfo {author} {\bibfnamefont {H.}~\bibnamefont {Berger}},
  \bibinfo {author} {\bibfnamefont {I.}~\bibnamefont {Vobornik}}, \bibinfo
  {author} {\bibfnamefont {G.}~\bibnamefont {Margaritondo}}, \bibinfo {author}
  {\bibfnamefont {L.}~\bibnamefont {Forro}}, \bibinfo {author} {\bibfnamefont
  {C.}~\bibnamefont {Beeli}}, \bibinfo {author} {\bibfnamefont
  {M.}~\bibnamefont {Onellion}}, \bibinfo {author} {\bibfnamefont
  {G.}~\bibnamefont {Panaccione}}, \bibinfo {author} {\bibfnamefont
  {A.}~\bibnamefont {Taleb-Ibrahimi}}, \ and\ \bibinfo {author} {\bibfnamefont
  {M.}~\bibnamefont {Grioni}},\ }\href@noop {} {\bibfield  {journal} {\bibinfo
  {journal} {Phys. Rev. Lett.}\ }\textbf {\bibinfo {volume} {81}},\ \bibinfo
  {pages} {1058} (\bibinfo {year} {1998})}\BibitemShut {NoStop}%
\bibitem [{\citenamefont {Clerc}\ \emph {et~al.}(2006)\citenamefont {Clerc},
  \citenamefont {Battaglia}, \citenamefont {Bovet}, \citenamefont {Despont},
  \citenamefont {Monney}, \citenamefont {Cercellier}, \citenamefont {Garnier},\
  and\ \citenamefont {Aebi}}]{26}%
  \BibitemOpen
  \bibfield  {author} {\bibinfo {author} {\bibfnamefont {F.}~\bibnamefont
  {Clerc}}, \bibinfo {author} {\bibfnamefont {C.}~\bibnamefont {Battaglia}},
  \bibinfo {author} {\bibfnamefont {M.}~\bibnamefont {Bovet}}, \bibinfo
  {author} {\bibfnamefont {L.}~\bibnamefont {Despont}}, \bibinfo {author}
  {\bibfnamefont {C.}~\bibnamefont {Monney}}, \bibinfo {author} {\bibfnamefont
  {H.}~\bibnamefont {Cercellier}}, \bibinfo {author} {\bibfnamefont {M.~G.}\
  \bibnamefont {Garnier}}, \ and\ \bibinfo {author} {\bibfnamefont
  {P.}~\bibnamefont {Aebi}},\ }\href@noop {} {\bibfield  {journal} {\bibinfo
  {journal} {Phys. Rev. B}\ }\textbf {\bibinfo {volume} {74}},\ \bibinfo
  {pages} {155114} (\bibinfo {year} {2006})}\BibitemShut {NoStop}%
\bibitem [{\citenamefont {Ma}\ \emph {et~al.}(2016)\citenamefont {Ma},
  \citenamefont {Ye}, \citenamefont {Yu}, \citenamefont {Lu}, \citenamefont
  {Niu}, \citenamefont {Kim}, \citenamefont {Feng}, \citenamefont {Tomanek},
  \citenamefont {Son}, \citenamefont {Chen},\ and\ \citenamefont {Zhang}}]{32}%
  \BibitemOpen
  \bibfield  {author} {\bibinfo {author} {\bibfnamefont {L.}~\bibnamefont
  {Ma}}, \bibinfo {author} {\bibfnamefont {C.}~\bibnamefont {Ye}}, \bibinfo
  {author} {\bibfnamefont {Y.}~\bibnamefont {Yu}}, \bibinfo {author}
  {\bibfnamefont {X.~F.}\ \bibnamefont {Lu}}, \bibinfo {author} {\bibfnamefont
  {X.}~\bibnamefont {Niu}}, \bibinfo {author} {\bibfnamefont {S.}~\bibnamefont
  {Kim}}, \bibinfo {author} {\bibfnamefont {D.}~\bibnamefont {Feng}}, \bibinfo
  {author} {\bibfnamefont {D.}~\bibnamefont {Tomanek}}, \bibinfo {author}
  {\bibfnamefont {Y.-W.}\ \bibnamefont {Son}}, \bibinfo {author} {\bibfnamefont
  {X.~H.}\ \bibnamefont {Chen}}, \ and\ \bibinfo {author} {\bibfnamefont
  {Y.}~\bibnamefont {Zhang}},\ }\href@noop {} {\bibfield  {journal} {\bibinfo
  {journal} {Nat. Commun.}\ }\textbf {\bibinfo {volume} {7}},\ \bibinfo {pages}
  {10956} (\bibinfo {year} {2016})}\BibitemShut {NoStop}%
\bibitem [{\citenamefont {Cho}\ \emph {et~al.}(2016)\citenamefont {Cho},
  \citenamefont {Cheon}, \citenamefont {Kim}, \citenamefont {Lee},
  \citenamefont {Cho}, \citenamefont {Cheong},\ and\ \citenamefont
  {Yeom.}}]{33}%
  \BibitemOpen
  \bibfield  {author} {\bibinfo {author} {\bibfnamefont {D.}~\bibnamefont
  {Cho}}, \bibinfo {author} {\bibfnamefont {S.}~\bibnamefont {Cheon}}, \bibinfo
  {author} {\bibfnamefont {K.-S.}\ \bibnamefont {Kim}}, \bibinfo {author}
  {\bibfnamefont {S.-H.}\ \bibnamefont {Lee}}, \bibinfo {author} {\bibfnamefont
  {Y.-H.}\ \bibnamefont {Cho}}, \bibinfo {author} {\bibfnamefont {S.-W.}\
  \bibnamefont {Cheong}}, \ and\ \bibinfo {author} {\bibfnamefont {H.~W.}\
  \bibnamefont {Yeom.}},\ }\href@noop {} {\bibfield  {journal} {\bibinfo
  {journal} {Nat. Commun.}\ }\textbf {\bibinfo {volume} {7}},\ \bibinfo {pages}
  {10453} (\bibinfo {year} {2016})}\BibitemShut {NoStop}%
\bibitem [{\citenamefont {Perfetti}\ \emph {et~al.}(2003)\citenamefont
  {Perfetti}, \citenamefont {Georges}, \citenamefont {Florens}, \citenamefont
  {Biermann}, \citenamefont {Mitrovic}, \citenamefont {Berger}, \citenamefont
  {Tomm}, \citenamefont {Hochst},\ and\ \citenamefont {Grioni}}]{46}%
  \BibitemOpen
  \bibfield  {author} {\bibinfo {author} {\bibfnamefont {L.}~\bibnamefont
  {Perfetti}}, \bibinfo {author} {\bibfnamefont {A.}~\bibnamefont {Georges}},
  \bibinfo {author} {\bibfnamefont {S.}~\bibnamefont {Florens}}, \bibinfo
  {author} {\bibfnamefont {S.}~\bibnamefont {Biermann}}, \bibinfo {author}
  {\bibfnamefont {S.}~\bibnamefont {Mitrovic}}, \bibinfo {author}
  {\bibfnamefont {H.}~\bibnamefont {Berger}}, \bibinfo {author} {\bibfnamefont
  {Y.}~\bibnamefont {Tomm}}, \bibinfo {author} {\bibfnamefont {H.}~\bibnamefont
  {Hochst}}, \ and\ \bibinfo {author} {\bibfnamefont {M.}~\bibnamefont
  {Grioni}},\ }\href@noop {} {\bibfield  {journal} {\bibinfo  {journal} {Phys.
  Rev. Lett.}\ }\textbf {\bibinfo {volume} {90}},\ \bibinfo {pages} {166401}
  (\bibinfo {year} {2003})}\BibitemShut {NoStop}%
\bibitem [{\citenamefont {Perfetti}\ \emph {et~al.}(2008)\citenamefont
  {Perfetti}, \citenamefont {Loukakos}, \citenamefont {Lisowski}, \citenamefont
  {Bovensiepen}, \citenamefont {Wolf}, \citenamefont {Berger}, \citenamefont
  {Biermann},\ and\ \citenamefont {Georges}}]{47}%
  \BibitemOpen
  \bibfield  {author} {\bibinfo {author} {\bibfnamefont {L.}~\bibnamefont
  {Perfetti}}, \bibinfo {author} {\bibfnamefont {P.~A.}\ \bibnamefont
  {Loukakos}}, \bibinfo {author} {\bibfnamefont {M.}~\bibnamefont {Lisowski}},
  \bibinfo {author} {\bibfnamefont {U.}~\bibnamefont {Bovensiepen}}, \bibinfo
  {author} {\bibfnamefont {M.}~\bibnamefont {Wolf}}, \bibinfo {author}
  {\bibfnamefont {H.}~\bibnamefont {Berger}}, \bibinfo {author} {\bibfnamefont
  {S.}~\bibnamefont {Biermann}}, \ and\ \bibinfo {author} {\bibfnamefont
  {A.}~\bibnamefont {Georges}},\ }\href@noop {} {\bibfield  {journal} {\bibinfo
   {journal} {New J. Phys.}\ }\textbf {\bibinfo {volume} {10}},\ \bibinfo
  {pages} {053019} (\bibinfo {year} {2008})}\BibitemShut {NoStop}%
\bibitem [{\citenamefont {Freericks}\ \emph {et~al.}(2009)\citenamefont
  {Freericks}, \citenamefont {Krishnamurthy}, \citenamefont {Ge}, \citenamefont
  {Liu},\ and\ \citenamefont {Pruschke}}]{48}%
  \BibitemOpen
  \bibfield  {author} {\bibinfo {author} {\bibfnamefont {J.~K.}\ \bibnamefont
  {Freericks}}, \bibinfo {author} {\bibfnamefont {H.~R.}\ \bibnamefont
  {Krishnamurthy}}, \bibinfo {author} {\bibfnamefont {Y.}~\bibnamefont {Ge}},
  \bibinfo {author} {\bibfnamefont {A.~Y.}\ \bibnamefont {Liu}}, \ and\
  \bibinfo {author} {\bibfnamefont {T.}~\bibnamefont {Pruschke}},\ }\href@noop
  {} {\bibfield  {journal} {\bibinfo  {journal} {Phys. Status. Solidi. B}\
  }\textbf {\bibinfo {volume} {246}},\ \bibinfo {pages} {948} (\bibinfo {year}
  {2009})}\BibitemShut {NoStop}%
\end{thebibliography}
\end{document}